\documentclass[a4paper,usenatbib]{mn2e}
\usepackage{hyperref,graphicx}
\usepackage{rotating}
\usepackage{amsmath,amssymb}
\usepackage{multirow}
\usepackage{setspace}
\usepackage{longtable}

\hypersetup{colorlinks=true, citecolor=blue, linkcolor= magenta}

\newcommand{\apj}{ApJ}
\newcommand{\apjl}{ApJ}
\newcommand{\apjs}{ApJS}

\newcommand{\mnras}{MNRAS}
\newcommand{\nat}{Nature}
\newcommand{\nar}{New Astronomy Reviews}
\newcommand{\aap}{A{\&}A}
\newcommand{\araa}{ARA{\&}A}
\newcommand{\aapr}{A{\&}ARv}

\newcommand{\pasj}{PASJ}

\newcommand{\aql}{Aql~X$-$1}
\newcommand{\aqls}{Aql~X$-$1~}

\title[Classification and Spectral Evolution of Outbursts of \aql]
{Classification and Spectral Evolution of Outbursts of \aql}

\author[C.\ G\"{u}ng\"{o}r, T.\ G\"{u}ver, \& K.\ Y.\ Ek\c{s}i]{C.\      G\"{u}ng\"{o}r$^{1}$,
  T.\ G\"{u}ver$^{2}$, \& K.\ Y.\ Ek\c{s}i$^{1}$ \\  
  $^{1}$Istanbul Technical University,
  Faculty  of Science  and  Letters,  Physics Engineering  Department,
  34469,  Istanbul, Turkey  \\ $^{2}$Istanbul University, Science Faculty, Department of Astronomy and Space Sciences, Beyaz{\i}t, 34119, Istanbul, Turkey}  

\begin{document}
\date{}
\pagerange{\pageref{firstpage}--\pageref{lastpage}} \pubyear{2013}
\maketitle

\begin{abstract}

We present a  broad classification of all outbursts  detected with the
All-Sky Monitor  (ASM) on the  Rossi X-Ray Timing Explorer  (RXTE) and
the Monitor of All Sky X-Ray Image (MAXI) of \aql.  We identify three types of
outbursts; \emph{long-high},  \emph{medium-low}, and \emph{short-low},
based on the duration and maximum  flux.  We analyse the trends in the
``phase-space''  of flux-derivative  versus  flux  to demonstrate  the
differences in the three identified outburst types.  We  present a
spectral  analysis  of the  observations  of  \aqls performed  by  the
Proportional  Counter Array  (PCA)  onboard RXTE  during  the 2000  and
2011 outbursts of the long-high class  and the 2010 outburst of the
medium-low  class.   We  model  the  source  spectrum  with  a  hybrid
thermal/non-thermal hot plasma emission  model (EQPAIR in XSPEC, Coppi
2000) together  with a  Gaussian component to  model the  Fe K$\alpha$
emission line.   We construct time  histories of the source  flux, the
optical  depth of  the corona  ($\tau$), the  seed photon  temperature
($kT_{\rm bb}$)  and the hard state  compactness ($l_{\rm h}$)
for these  three outbursts.  We  show that the physical  parameters of
either  classes reach the same values throughout the outbursts, the
only difference being the maximum flux.  We discuss
our results  in the terms of  modes of interaction of the star with
the disc and size of the disc kept hot by irradiation. 
We conclude that irradiation is the dominant physical process leading 
to the different classes of outbursts.
\end{abstract}

\begin{keywords}
X-rays: binaries X-rays:individual \aqls
\end{keywords}

\label{firstpage}

\section{INTRODUCTION}

\aqls is  a low  mass X-ray  binary in which  a neutron  star accretes
matter from  an accretion  disc which  is supplied  by the  Roche lobe
filling low mass companion.  \aqls is  also classified as a soft
X-ray  transient \citep[SXTs,  see][for  a review]{cam+98rev}  system;
most of the time it is in a quiescent state with a luminosity of
$L_{\rm   X}  \approx   10^{33}$~erg~s$^{-1}$  \citep{verb+94}   while
occasionally  it exhibits  outbursts which  at the  peak can  reach to
$L_{\rm X}  \approx 10^{37}$~erg~s$^{-1}$ resulting from  the enhanced
accretion  rate,  $\dot{M}$,  in  the disc.   According  to  the  disc
instability   model   \citep[DIM;   see][for  a   review]{las01}   the
viscous-thermal instability  in the disc \citep{vanparadijs96}  is the
cause of the enhanced accretion.

As with  all transient systems,  the X-ray morphology of  outbursts of
\aqls show a wide range  of lightcurve patterns.  A typical lightcurve
pattern  is described  as  a fast-rise-exponential-decay  \citep[FRED;
][]{che+97}.  A FRED lightcurve can be reproduced with the DIM only if
the irradiation of the outer parts of  the disc by the X-rays from the
central source is taken into account \citep{kin98}.  Apart from events
that show  FRED like behavior  and its varieties  there is also  a low
intensity state \citep[LIS; ][]{wac+02} in  which the optical and near
IR  emission flux  is above  quiescence  but the  lightcurve does  not
follow the FRED trend and can last longer. 
What  leads to such distinct outburst lightcurves  is not very
well understood.

A clue to the diversity of the outburst lightcurve morphology could be
the variety  of spectral states  the transient systems  exhibit. An SXT 
in the rising phase of the outburst would usually  enter the high/soft  
(HS) state from the  low/hard (LH) state and  return to the  LH state 
during  the decay of  the outburst \citep[see][for a review of 
spectral states in black hole systems]{rem06}. Spectral states of SXTs 
in which the accreting objects are neutron stars are similar to those 
in black hole systems. This similarity is not well understood given that the neutron 
stars have hard surfaces where the energy of the accreting matter would
be thermalized whereas the black holes have event horizons through which 
the energy could be advected with the accretion flow.
\aqls  is classified  as  an  atoll source  \citep{vanderklis94,has89}
according to the tracks it follows  on the X-ray color-color and
hardness-intensity  diagrams, and  spectral variations  it shows.   
In this case  HS and LH  states roughly  correspond to banana  and island
states \citep{has89}, respectively.

The  HS  state  is  associated  with  a  standard  geometrically  thin
optically thick  Keplerian disc  \citep{sha73} while  the LH  state is
associated  with  a  geometrically   thick  optically  thin  advection
dominated disc \citep{ich77,nar94} and a corona at which the soft seed
photons  from  the disc  blackbody  are  up-scattered by  relativistic
electrons.  In order  to address the low luminosity in  the hard state
it is also  necessary to introduce the truncation of  the inner region
of the geometrically thin accretion disk.

The different  modes of interaction  between the magnetosphere  of the
neutron  star and  the  surrounding  disc,  that  take place  at
different  accretion rates,  which  changes by  three orders  of
magnitude during  an outburst, may also  have a role in the diversity of the
lightcurve morphology.  During the decay  phase in which the mass flux
is decreasing, the inner radius of  the disk moves outwards.  Once the
inner  radius   is  pushed  beyond  the   corotation  radius,  $R_{\rm
  c}=(GM/\Omega^2)^{1/3}$, at which Keplerian  angular velocity in the
disk equals the angular velocity  of the magnetosphere $\Omega$, it is
expected that accretion onto the  star is inhibited by the centrifugal
barrier  and the  system  is said  to be  in  the ``propeller''  stage
\citep{ill75,dav79,dav81,wan85}.   \aqls  showed   a  rapid  decay  of
luminosity  accompanied by  an abrupt  spectral transition  during the
decline of its 1997 outburst which was interpreted as the onset of the
``propeller'' stage \citep{cam+98,zha+98pro}.

\aqls  is one  of the  most active  SXTs making  it possibly  the most
suitable system for classifying outbursts, investigating the interplay
between  irradiation,  transition  to  propeller  stage  and  spectral
transitions.  In  \S\ref{class}, we present a  broad classification of
outbursts  based  on  the  durations   and  the  maximum  fluxes.   In
\S\ref{obs},  we  explain the  details  of  observation and  analyzing
procedure,  and   provide  the  results  of   observational  analysis.
In \S\ref{discuss}, we discuss  the evolution of the physical
parameters which we  obtained and we associate them  with the outburst
types we classified. Finally, in \S\ref{conclude} we present our conclusion. 

\section{Classification of outbursts of AQL X$-$1}
\label{class}

\begin{figure}
\centering
  \includegraphics[angle=0, scale=0.5]{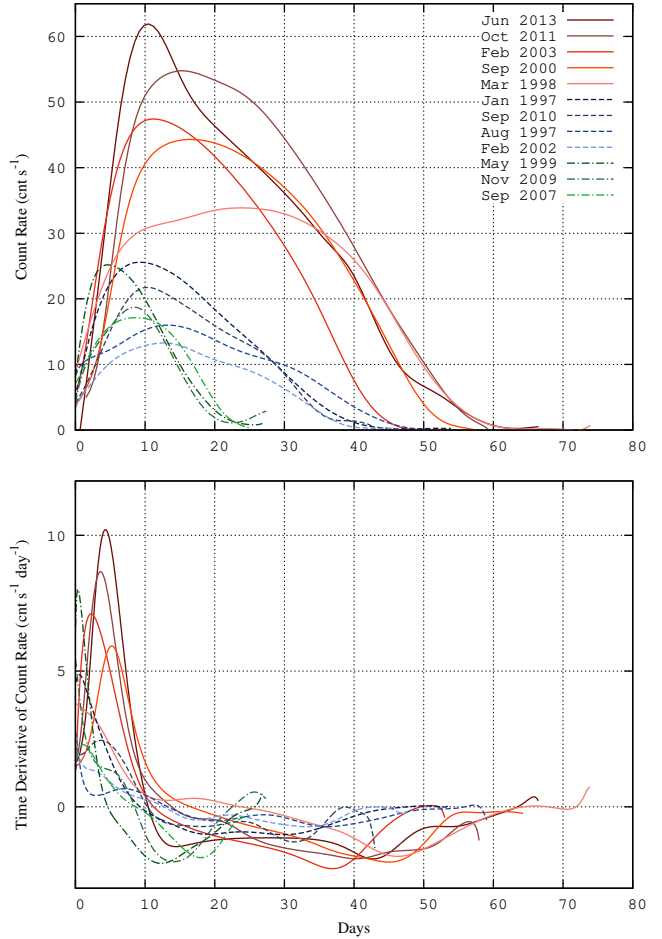}
     \caption{ASM and MAXI lightcurves  of outbursts (upper panel) and
       their  time derivative.   The  data is  smoothed with  B\'ezier
       process  for visualization.   Three types  of outbursts  can be
       discriminated: (i) The long-high outbursts  shown  with solid
       lines (red).  These events  last about 50-60  days and  reach a
       count rate of 37-61  cnt/s.(ii) The medium-low outbursts, shown
       with dashed  lines (blue),  last about 40-50  days and  reach a
       count  rate of  13-25  cnt/s.  (iii)  The short-low  outbursts,
       shown  with dashed-dot lines  (green), last  about 20  days and
       reach a count rate of 17-25 cnt/s.}
         \label{asm_plot}
\end{figure}

We present a  broad classification of the outbursts of  \aqls based on
the duration and the maximum  flux. A different classification for the
outbursts  of this  object  was recently  presented by  \citet{asai+13}
based on  the spectral transitions \citep[see  also][]{mai08, cam+13}.
The classification  we present may not be an  alternative to this 
as will be discussed in \S \ref{trans}.
In  Figure~\ref{asm_plot} (upper panel), we
show all outbursts  as observed by ASM \citep{lev+96}  aboard the RXTE
\citep{jah+96}; the 2011 and the 2013 outbursts observed by MAXI
\citep{mat+09}.  For each instrument we  used the daily average values
in the 2$-$10  keV and 2$-$20~keV range,  respectively.  We calibrated
the data  sets from the  two instruments  by comparing the  peak count
rate  of the  2009 and  2010 outbursts,  which were  observed by  both
detectors.  Using  the near-simultaneous  detections, we  determined a
conversion factor of $\approx$22.  The beginning of the outbursts were
determined as the  point at which the count-rate reaches  5 cnt/s.  
To see the 
trends in the lightcurves more clearly we
have smoothed the data sets using the B\'ezier spline method.

The  upper  panel of  Figure  \ref{asm_plot}  shows the  evolution  of
count-rate in all data sets.  It is clearly seen that 2013 outburst is
the brightest outburst  of \aqls in the data set,  as was suggested by
\citet{guver+13}.  It is also  seen that there are three types of
outbursts.   The \emph{long-high}  outbursts have  long duration  with
50-60 days and  are luminous, reaching a maximum flux  of 37-61 cnt/s.
The \emph{medium-low} outbursts last for  40-50 days, reaching a maximum flux
of 13-25 cnt/s.  The \emph{short-low} outbursts last for approximately
20  days reaching  a  maximum flux  of 17-25  cnt/s.   Note that  this
classification does  not include  outbursts corresponding to  LIS that
last even longer than the long-high outbursts at a LH state, but never
reaches to  count-rate of  5 cnt/s  level, which  is our  criterion to
calibrate initial times of all outbursts.

Conversion of  the data sets  to continuous curves by  B\'ezier spline
method allows us  to evaluate the time derivative of  the flux curves,
which we show in the lower panel of Figure \ref{asm_plot}. We see that
the rising stages of the bright outbursts consist of two sub-stages in
which  the  flux  change  rate  increases  and  then  decreases.   The
medium-low outbursts show  a more complex trend. It is  also seen that
the decay stage  also roughly consists of two sub-stages  in which the
flux decay rate  first speeds-up and then slows down.  The behavior at
the lowest flux levels are  much more complicated, showing variability
at a time-scale of days.

\begin{figure}
\centering
  \includegraphics[angle=0, scale=0.5]{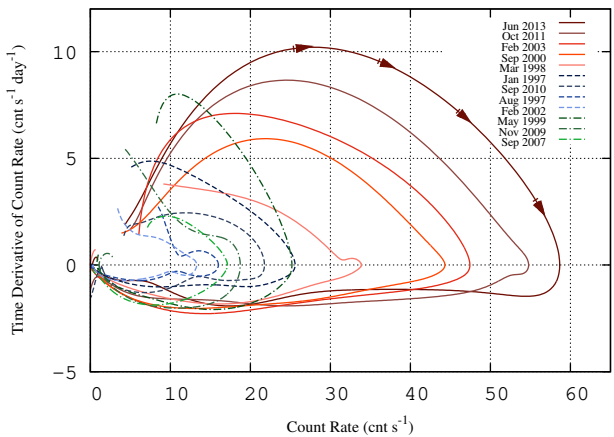}
     \caption{The  outbursts  in   the  counts/s  versus  counts/s/day
       ``phase  space''. The  arrows on  the curve  of June 2013 event
       shows  the direction  of  time.  Outbursts  start from the upper-left domain,
      move clockwise, pass  from  the $y=0$  at the peak of the outbursts,  and
       reach to the lower-left domain (quiescence).}
         \label{phase}
\end{figure}

Figure \ref{phase} shows  the evolution of the outbursts  in the ``phase
space''  of  flux-rate  versus  flux. An outburst  starts  from
the upper-left domain of the  plot, moves clockwise as shown with arrows  on the
outermost curve for the June 2013 and ends
near the origin.


\section{Observation and data analysis}
\label{obs}

In  order to  better  characterize the  distinction between  different
types of  outbursts we used  the monitoring observations  performed by
RXTE/PCA.   We  analysed  the  2000,  2010, and  2011  outbursts  of
\aql.  Based  on the  above  mentioned  classification,  the 2000  and
2011 outbursts  are members of  the long-high class  and the
2010 outburst is a member of the medium-low class.

\begin{figure}
\centering
  \includegraphics[angle=0, scale=0.48]{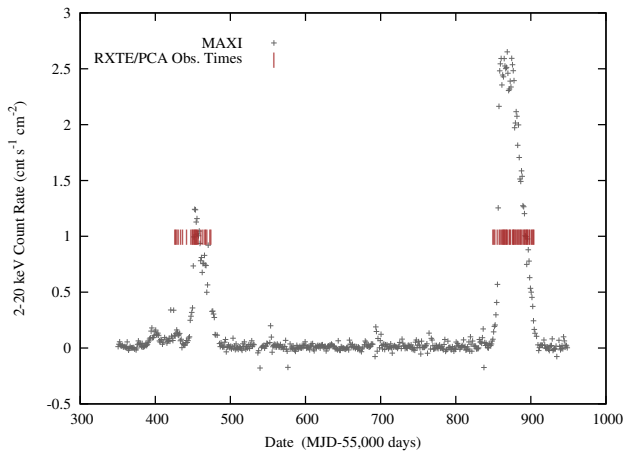}
     \caption{The  long-term evolution  of \aqls  as observed  by MAXI
       between  June   2010  to  January  2012.    Dates  of  RXTE/PCA
       observations are marked with vertical bars.}
         \label{maxi_plot}
\end{figure}

The times of  RXTE observations during 2010 and  2011 outbursts are
shown on the MAXI lightcurve in Figure \ref{maxi_plot}.  It is clearly
visible  that the  2011 outburst  is much  more luminous  than 2010
outburst  and  the  RXTE  pointings  in 2011  covers  the  outburst
completely, including both the rise and decay phases. We  also focused on
the  2000 outburst  as  a comparison,  since  it is  one  of the  most
luminous outbursts with detailed RXTE observations.

As \aqls is a bright source, using  one of the four PCUs is sufficient
to produce high S/N spectrum.  We used PCU2, which has been calibrated
the  best and  has been  continuously working.   We analysed  the data
using HEAsoft v6.11 and created  response matrices separately for each
data set  using the  PCARSP version 11.7.   We created  the background
spectra   using   the  latest   module   file   for  bright   sources,
\emph{pca$\_$bkgd$\_$cmbrightvle$\_$eMv20051128.mdl}.

We used Xspec v12.7.0 for the spectral analysis. After subtracting the
background from  the original  data, we fit  3$-$30 keV  energy range,
retaining the band at which the PCA detector is the most sensitive. We
added a systematic error of 1\%.   During the 2010 and 2011 outbursts,
we   determined   2  and   4   Type-I   thermonuclear  X-ray   bursts,
respectively. We  ignored the events detected  100 seconds before
  the start and 100 seconds after the end of these bursts.  Following
\citet{mac03spec}  we  fitted  the  spectrum  with  the  EQPAIR  model
\citep{ref_eqpair}, which  is a  hybrid comptonization  model bridging
the   purely  Maxwellian   and   purely   power-law  particle   energy
distributions, responsible for the emission, in a self-consistent way.
EQPAIR uses the ratio of relative  heating and cooling of electrons on
account of  radiative processes  and Coulomb interactions  for solving
the electron  distribution.
Many other models have been employed in the literature for
fitting the X-ray spectra of LMXBs \citep[see  also][]{rai+11, mai04}. 
Statistically, all these models fit the X-ray spectra  as  successfully as EQPAIR.
EQPAIR has  been used  for fitting  X-ray spectra  of disc  flows with
corona in  LMXBs with  black hole  components \citep{gar08}.   As such
EQPAIR can  not address  how much the neutron  star contributes to the thermal
 and non-thermal emission from these sources. Our  choice of  EQPAIR  for
modeling the spectra is motivated  by the purpose of investigating the
role of the  parameters of the corona in the  irradiation of the outer
disc.  Using EQPAIR also  allows us  to compare our  results with
earlier studies \citep{mac03spec}. Since EQPAIR  is a hybrid model for
both plasma, it  is able to fit  all the energy range  between 3.0 keV
and 30.0 keV, with the  addition of a statistically necessary Gaussian
component.

\begin{figure*}
\centering
  \includegraphics[angle=0, scale=0.5]{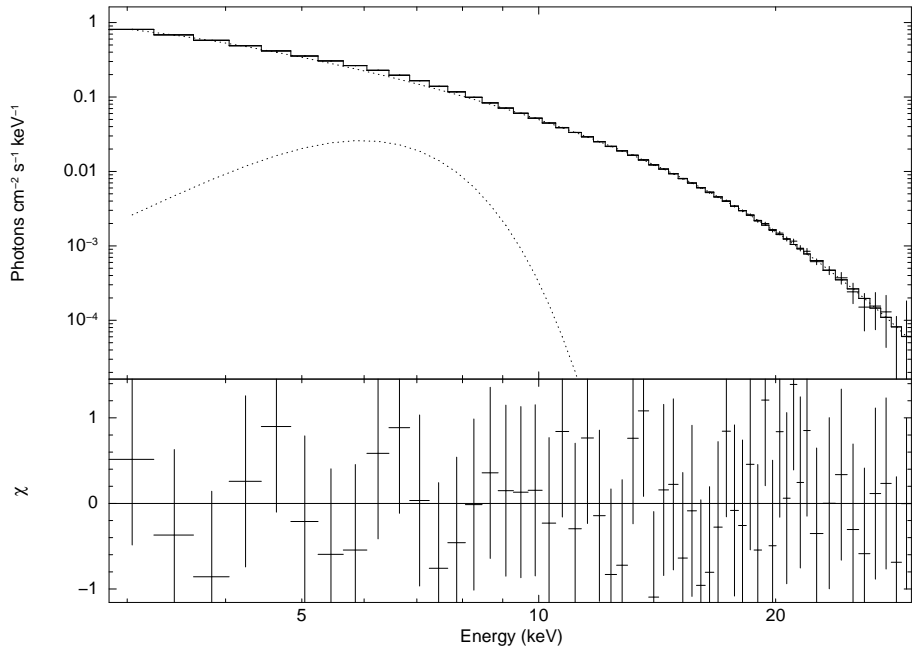}
     \caption{Upper   panel  shows   the   X-ray   spectrum  (Obs   ID
       96440-01-03-04) and the best fit model between 3.0 keV and 30.0
       keV  together with  the best  fit EQPAIR  model and  a Gaussian
       line.  Lower panel shows the residuals in terms of sigma.}
         \label{fit_pgplot}
\end{figure*}

We fixed the neutral hydrogen column density at $N_{\rm H}~=~3.4\times
10^{21}~{\rm cm}^{-2}$ \citep{mac03spec}.  The soft photon compactness
($l_{\rm  bb}$)  and  the  fraction of  power  supplied  to  energetic
particles which goes into  accelerating non-thermal particles ($l_{\rm
  nt}/l_{\rm  h}$) were  frozen at  1.0 and  0.0, respectively.   This
allows  us  to  determine  $l_{\rm  h}$  in  terms  of  $l_{\rm  bb}$. 
\citet{wel00}  shows that  the orbital  inclination should  be greater
than 36  degree. The inclination  angle was  chosen as 60  degrees for
simplicity.  All  the other  parameters of EQPAIR  model, were  set to
default  values  and  kept  frozen  during  the  fit
  \citep[following][]{mac03spec}.  After  we reached  the best fit  parameters of
EQPAIR model with  minimum $\chi^2$, we added a  Gaussian component to
take  into account  the  systematic  residuals due  to  the iron  line
between  5.8  keV  and  7.0~keV   for  each  observation.   In  Figure
\ref{fit_pgplot}, we show an example  X-ray spectrum of \aqls obtained
during the peak  of the 2011 outburst (Obs ID  96440-01-03-04) and the
best fit EQPAIR model with a Gaussian line.


For all the X-ray spectra  of \aqls obtained by RXTE/PCA observations,
we  followed the  procedure described  above.  Overall,  for the  2000
outburst we  have 55 pointings,  when the  source was in  the outburst
phase.  The  2010 outburst  was covered with  24 pointings.   Seven of
these  observations  were  obtained  when  the  source  was  still  in
quiescent phase, which allows us to better characterize how parameters
evolve as the source advances  from quiescence to outburst. During the
2011 outburst,  we have 43 data  sets that completely cover  both, the
rise and decay phases.

\begin{figure*}
\centering
  \includegraphics[angle=0, height=0.3\textheight, width=0.33\textwidth]{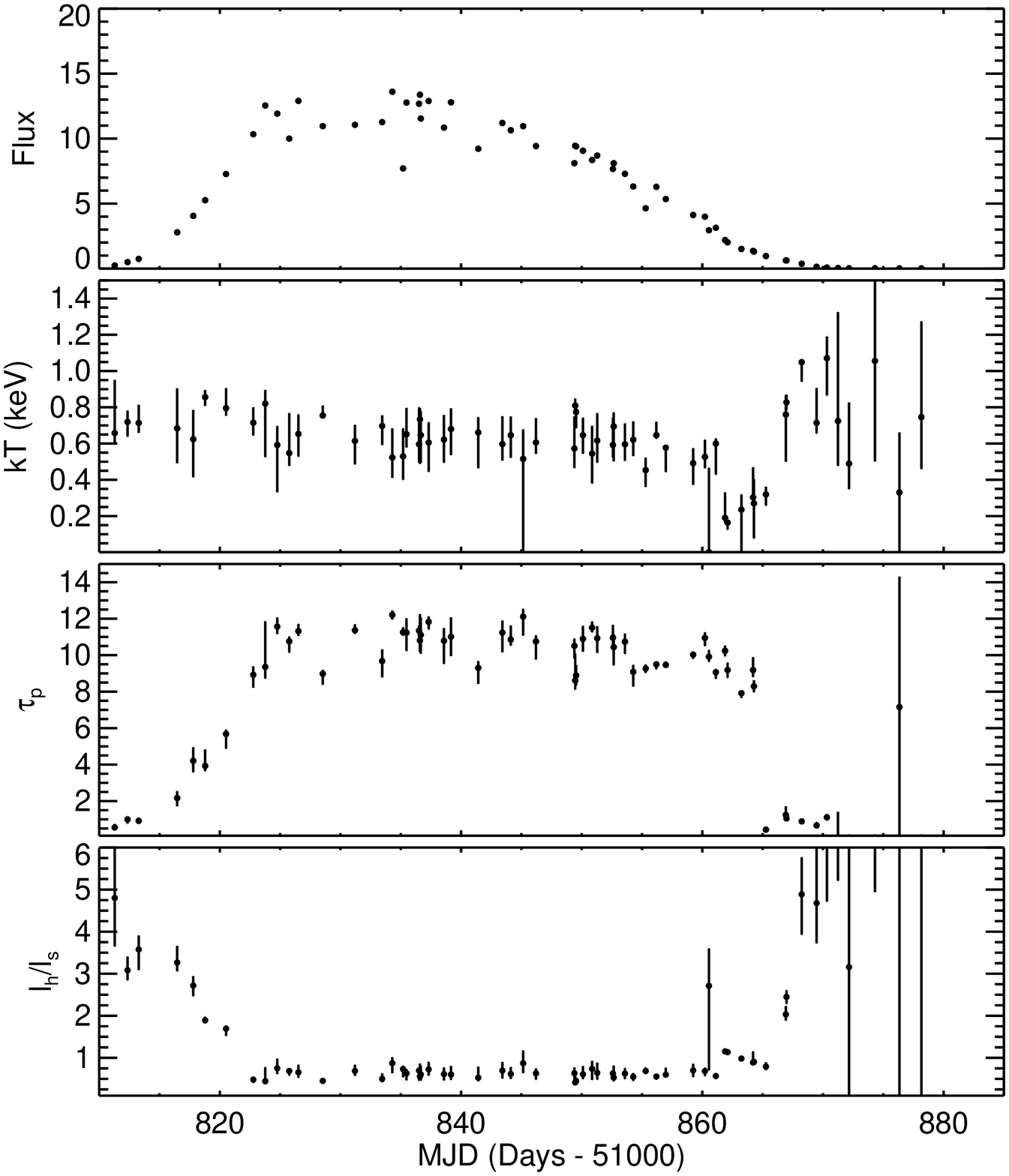}
  \includegraphics[angle=0, height=0.3\textheight, width=0.33\textwidth]{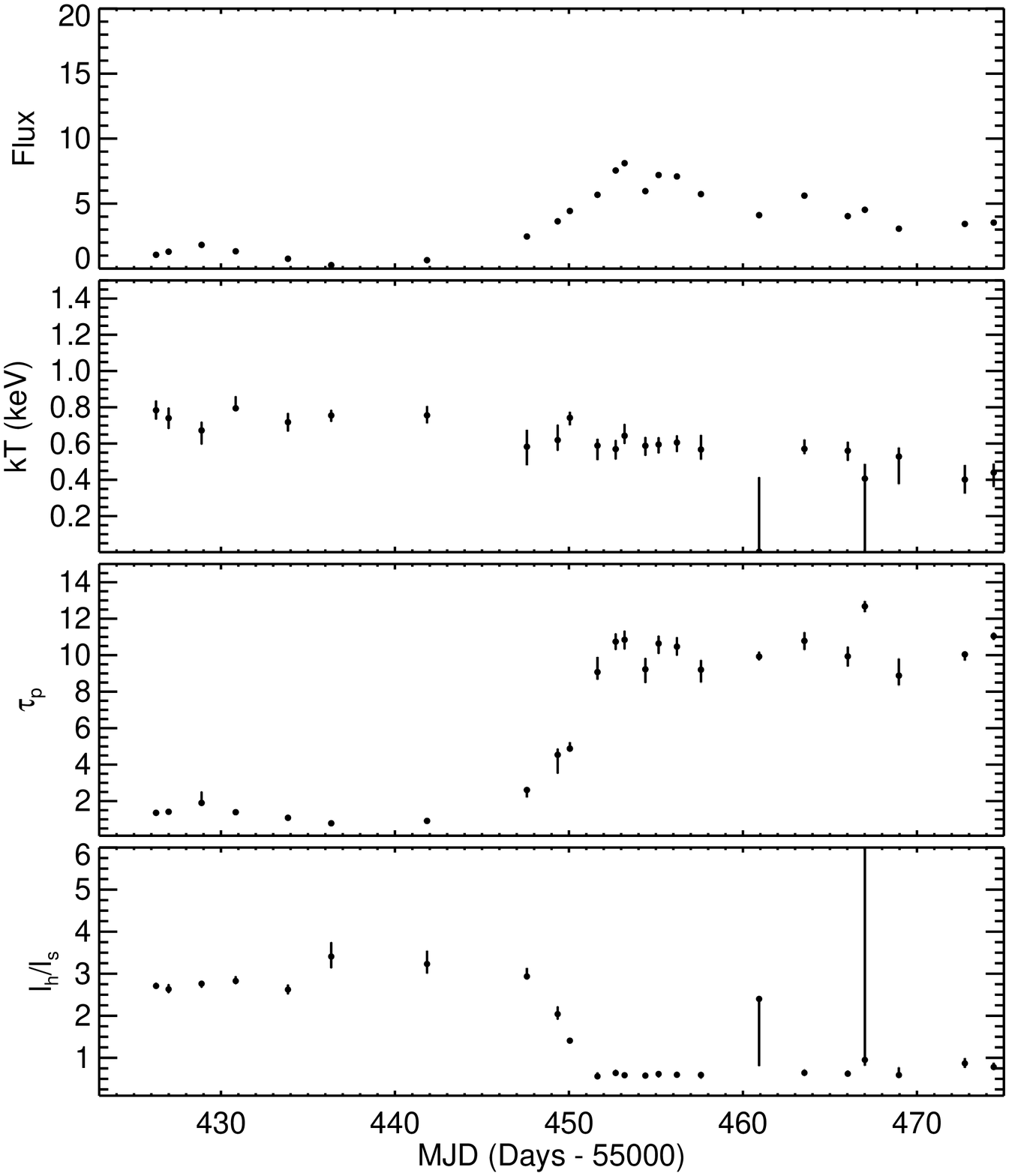}
  \includegraphics[angle=0, height=0.3\textheight, width=0.33\textwidth]{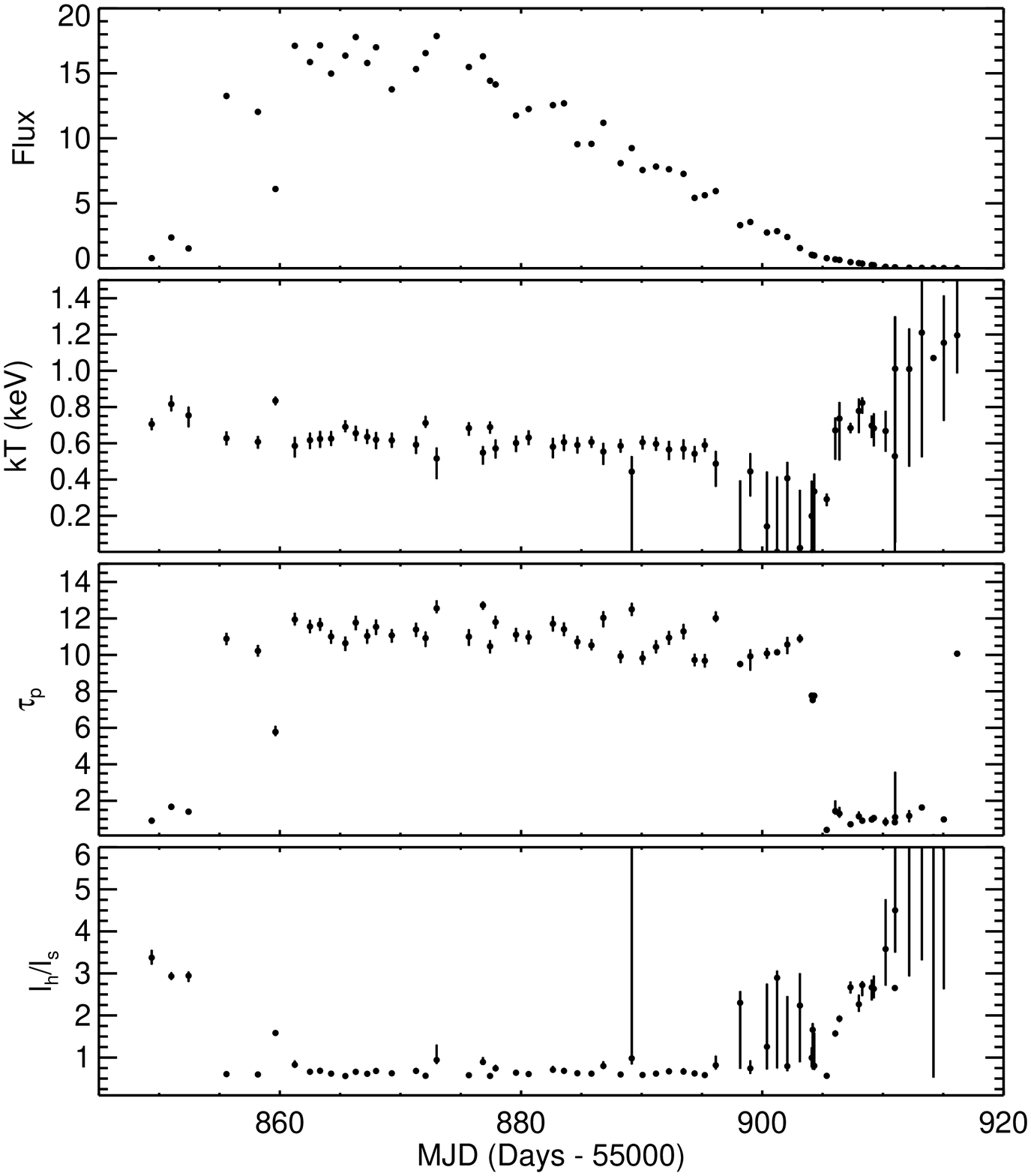}  
     \caption{Spectral   evolution   of    \aqls   during   the   2000 (left panel), 
     2010 (middle panel) and 2011 (right panel)
       outbursts.  From upper to  lower, panels  show the  evolution of
       unabsorbed total flux, temperature of the seed photons, optical
       depth,   and   the  ratio   of   hard   to  soft   compactness,
       respectively. All fluxes are given in units of $10^{-9}$~erg~s$^{-1}$~cm$^{-2}$.}
         \label{plot_all}
\end{figure*}

The  free parameters  of EQPAIR  model are  the optical  depth of  the
corona ($\tau$),  the seed photon temperature  ($k_{\rm B}T_{\rm bb}$)
and the  hard state compactness ($l_{\rm  h}$).  Figure \ref{plot_all}
shows the  time evolution  of these parameters  during 2000,  2010 and
2011 outbursts. The  spectral parameters evolve in a  very similar way
in each outburst: {\it (i)} The  optical depth of the corona increases
during the rise  and becomes stable until the end  of the decay.  {\it
  (ii)}  The   hard  state   compactness  decreases   to  $\approx$0.5
immediately after the onset of outbursts.  {\it (iii)} We see that the
seed photon  temperature slightly decreases as  the outbursts proceed.
We note  however that this is  the least constrained parameter  in our
model  because  the inferred  values  are  too  low for  the  RXTE/PCA
sensitivity range,  during the outburst.   The best fit  parameters of
EQPAIR model and Gaussian component  including the errors are shown in
Table   \ref{tab:par.2000},   Table   \ref{tab:par.2010}   and   Table
\ref{tab:par.2011-12}.

\section{Discussion}
\label{discuss}

We  consider three  different  causes for  the  diversity of  outburst
morphology  presented in  \S \ref{class}:  (i) A  longer waiting  time
before an outburst might lead  to larger amount of matter accumulation
in the  disc resulting in long-high  type outbursts; (ii) it  might be
that only  during the \emph{long-high}  outbursts the inner  radius of
the  disc  penetrates the  corotation  radius  leading to  substantial
accretion  while  the   \emph{medium-low}  and  the \emph{short-low}  both
proceed  by partial  accretion during  the spin-down  regime; (iii)  a
different  portion of  the  disc  is involved  in  different types  of
outbursts  probably   because  of  different   irradiation  geometries
involved. In  the following we  discuss these three options  in detail
 and discuss  the implications  of  the results  of our  spectral
  analysis.

\subsection{Is the quiescent stage duration related to the different outburst classes?}

One naturally expects that a longer waiting time between two outbursts
might  lead to a larger  amount of  matter accumulation  in the  disc
resulting with a more luminous  outburst.  This possibility was considered
by \citet{cam+13} (see their fig.\ 6) who checked whether the peak flux
of outbursts  are correlated with  the preceding waiting  times before
the outbursts. They found only a very weak correlation indicating that
the  duration of  the quiescent  stage is  not the  main cause  of the
difference between the long-high, medium-low and short-low classes.

\subsection{Transition to the propeller stage as the cause of the rapid decay stage}
\label{trans}

The rapid  decay of luminosity  during the  decline of  the 1997
outburst of  \aqls was interpreted  as the onset of  the ``propeller''
stage \citep{cam+98,zha+98pro}. A similar rapid decline stage is shown
also by the first discovered  AMXP \citep{wij98} SAX~J1808.4$-$3658 in
several  of its  outbursts.  This  was associated  with transition  to
propeller stage by  \citet{gil+98}.  The 1997 outburst  of \aqls which
led to the propeller interpretation is of the low-medium class. It can
be seen from  Figure \ref{asm_plot} that, compared to  the 1997 event,
the 1998  outburst shows  a much more  prominent transition  from slow
decay to rapid  decay at a count-rate level  of $\approx$27.  Assuming
that this flux  level corresponds to the critical mass  flow rate that
places the  inner radius on  the corotation  radius, one is  forced to
conclude that the system remains in the spin-down stage throughout the
medium-low and short-low  outbursts, accreting only a  fraction of the
inflowing  mass because  the  maximum count-rate  for these  outbursts
remain below $\approx$27.   Within this scenario the  inner radius can
penetrate  the corotation  radius  only for  the brightest  outbursts,
which possibly  could address why  many outbursts are stunted  and why
pulsations are elusive.

In the case of SAX J1808.4$-$3658 (but not \aql ) the X-ray pulsations
continue   to   be   detected   even  in   the   rapid   decay   stage
\citep{psa99}. In  order to  address this issue  \citet{rap+04} argued
that it is  not energetically possible for the  magnetosphere to eject
matter from the system as  the magnetosphere expands to the corotation
radius with the  decaying mass flux. They presented  disk solutions in
which the  inner radius of the  disk remains on the  corotation radius
for  a  large  range  of  mass flux  such  that  accretion  and  X-ray
pulsations continue.   \citet{ibr09} concluded from the  variations of
the pulse profiles that  the size of the hot spot  on the neutron star
changes throughout  the outburst which  then indicates that  the inner
radius  of the  disc does  recede during  the decay  of the  outburst.
\citet{eks11} associated  the rapid decay stage  of SAX J1808.4$-$3658
with  a partial  accretion regime  \citep{men99,ust06,rom04} in  which
matter at the  disc mid-plane is centrifugally  inhibited while matter
vertically away  from the  mid-plane can  accrete.  This  requires the
inner region  of the  disc to  be geometrically  thick as  required by
radiatively inefficient disc models that address the LH state.

The onset of  a propeller during an outburst may  affect the number of
thermonuclear  X-ray bursts  occurring  during an  outburst.  To test  the
propeller  scenario, we  therefore  investigated  the observed  Type-I
X-ray  bursts  during  different  outbursts.  Using  the  X-ray  burst
catalogue of \citet{gal+08}  we identified the number  of X-ray bursts
detected during a given outburst. Although such an analysis is subject
to selection  effects due to the  varying coverage of the  RXTE/PCA of
individual outbursts, we  found that a similar number  of X-ray bursts
and  photospheric  radius expansion  events  have  been observed  from
different  types  of outbursts.   For  example,  during the  short-low
outburst observed  in 1999,  two photospheric radius  expansion events
were  observed together  with one  Type-I  burst.  On  the other  hand
during the 2000 outburst, which is  a long-high event according to our
classification, 4 Type-I X-ray bursts were  observed of which one is a
photopsheric radius expansion event.  A similar trend is also observed
during the  two outbursts  observed in  1997 and  2002, which  are all
medium-low type outbursts.  During these  outbursts 1, 4, and 5 type-I
X-ray bursts have been observed with one photospheric radius expansion
events  during each  of  the  1997 outbursts  and  2  during the  2002
outburst. We note  however that there are exceptions,  for example the
short-low outburst  that occurred in  1998 was observed  with RXTE/PCA
with  24 pointings.   Despite the  dense coverage  not a  single X-ray
burst was observed during the outburst from the source.  This seems to
support the argument discussed above  that there had been a transition
to the propeller stage during the 1998 outburst.

Recently, \citet{asai+13}  suggested transition to the  propeller stage
as  an explanation  for  the  rapid decay  in  the  lightcurves of  4U
1608$-$52,  \aqls and  XTE  J1701$-$462  discriminating transition  to
propeller  stage from  the  spectral transitions  by using  luminosity
dwell-time  distributions \citep[see  also][]{mat13}.   This does  not
mean that irradiation  is not present in the system,  but only that it
may  not be  the cause  of the  rapid decay  stage.  Presence  of jets
\citep{tud+09} in this system associated with transition from HS to LH
state may  also indicate  the role of  the ``propeller''  for ejecting
matter from the system (see e.g.\ \cite{rom+09,lii+12,ust+06,lov+99}).

A well known argument against the propeller interpretation is that not
only neutron star SXTs but some  black hole SXTs also show a ``brink''
in the  decay stages of their  outbursts implying a common  origin for
the rapid  decay \citep[e.g.\  ][]{jon+04bh}.  As  black holes  do not
have a magnetosphere, propeller mechanism can  not be at work in these
systems  and thus  can not  be the  common cause  for the  rapid decay
stage.  The presence  of similar spectral transitions  also suggests a
common mechanism disfavouring transition to  propeller as the cause of
the spectral transitions.

Yet  another  argument against  the  propeller  interpretation is  the
hysteresis effect: \citet{mac03hyst} showed that the LH to HS spectral
transition at the  initial rise of the burst occurs  at a luminosity 5
times  greater than  that  of the  transition  from HS  to  LH at  the
declining stage suggesting  that propeller mechanism alone  can not be
the driver of the spectral transitions as one would expect transitions
between accretion  and propeller stages  to be at a  single luminosity
for  a  certain  system \citep{dav73}.   More  recently,  \citet{yu07}
discovered, during the declining stage of the very faint 2001 outburst
of \aql,  a spectral transition  from LH  to HS state  suggesting that
mass flow rate,  $\dot{M}$, is not the sole  parameter determining the
luminosity of the  transition and mentioned the role of  change in the
mass flux.

These  problems  with  the  propeller  interpretation  can  partly  be
surpassed by suggesting that the  transition to propeller stage is not
the cause of the spectral transition  but a consequence of it.  If the
inner disc  makes a  transition to  a geometrically  thick radiatively
inefficient flow causing the transition to the LH state, the mass flux
and ram pressure suddenly drops as the flow becomes sub-keplerian; the
magnetosphere pushes the disc outwards which may then cause the system
to transit  to the propeller  stage if the  disc is pushed  beyond the
corotation radius.

\subsection{Irradiation as the parameter inducing different outburst durations and peak fluxes}

The difficulties  in interpreting the  rapid decline as  transition to
the  propeller  stage led  to  the  favored  view  that the  cause  of
transition  to the  rapid decay  stage is  related to  the outer  disc
disconnecting from the  inner disc by entering  the cool low-viscosity
state as this region is no longer irradiated \citep{pow+07}, the third
possibility we would like to discuss.

\begin{figure}
\centering
  \includegraphics[angle=0, scale=0.5]{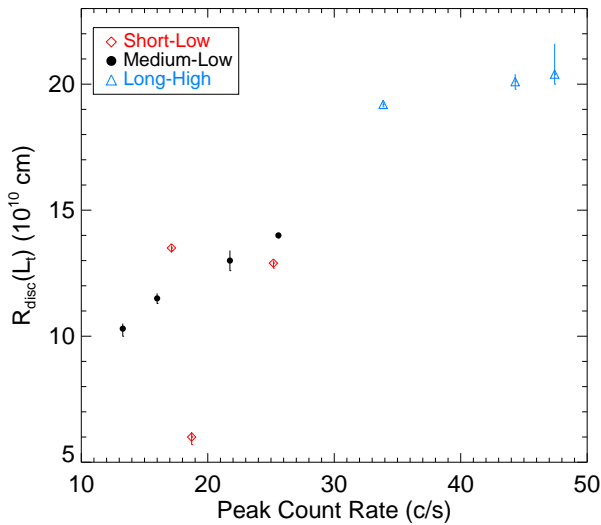}
     \caption{Outer  disc   radius  versus  the  peak   count-rate  of
       outbursts  of  \aql.  The  outer radii  of  the  outbursts  are
       estimated by  \citet{cam+13} via the furthest  point irradiated
       in the disc.  We see a linear correlation as  would be expected
       that a  larger outburst would  require a larger portion  of the
       disk  to be  involved in  the outburst.   Different classes  of
       outbursts are  shown with different symbols.  The long-high and
       the  medium-low outbursts  have  a tighter  correlation with  a
       different slope while short-low class shows no correlation.}
         \label{radius}
\end{figure}

Recently,  \citet{cam+13}  presented  an   in-depth  analysis  of  the
outbursts  observed from  \aqls  using the  irradiated  disc model  of
\citet{pow+07} and  fitted the  observed lightcurves.  This  brings in
two different  estimates of the outer  radius of the disc:  either via
the viscous  timescale at the outer  radius or via the  furthest point
that is irradiated  in the disc.  One naturally expects  that the peak
flux  would be  proportional  to the  size  of the  disc  kept hot  by
irradiation. In  Figure~\ref{radius}, we  present the  estimated outer
disc   radius   versus   the    peak   count-rate   we   identify   in
\S\ref{class}. The   outer  radius   values  are   taken  from
  \citet{cam+13}  who estimated  them based  on the  irradiation model
  \citep{pow+07}.   We obtained  corresponding  peak  fluxes from  the
  smoothed outbursts in \S~\ref{class}. We  find that the peak flux is
  roughly correlated with  the disk size.  As the maximum  size of the
  disc  is  determined by  the  Roche  lobe  radius, only  during  the
  long-high type outbursts the disc can have radii reaching that range
  while the  disc during  the medium-low  and the  short-low outbursts
  either does not fill the Roche  lobe radius, or that the outer parts
  of such  a disc  is not  sufficiently hot  to be  active and  is not
  involved  in the  outburst.   If the  latter is  the  case then  the
  correlation of the  peak flux with the disc size  is consistent with
  the  view  that larger  portion  of  the  disc  is involved  in  the
  long-high  type   of  outbursts   probably  due  to   the  different
  irradiation efficiency of the outer parts.

Although the  number of data for  each individual type of  outburst is
small,  it is  possible  to infer  from  Figure~\ref{radius} that  the
long-high and the medium-low outbursts have different slopes while the
short-low  outbursts show  a large  scatter  and does  not reveal  its
trend.  The smaller slope of  the long-high outbursts implies that for
such large  sized discs a  slight increase in  the disc size  leads to
much  higher peak  flux.  This is  because the  amount  of matter  and
angular momentum within a ring of definite width increase rapidly with
radial distance from the center. Thus a small change in the irradiation 
geometry may lead to a small change in the heated outer radius which 
then would lead to drastic changes in the outburst morphology.

\subsection{Implications of spectral analysis}

We have  considered the  possibility that  the difference  between the
three outburst  types of  the classification  in \S  \ref{class} could
originate from  different spectral  properties the disc  displays.  In
order  to  check  this  possibility  we  have  analyzed  the  spectral
evolution during  the 2000, 2010 and  2011 outbursts which are  of the
long-high type (2000 and 2011)  and medium-low type (2010).  With this
analysis our aim is to better  characterize these events and to better
understand    their    differences.    Using    the    EQPAIR    model
\citep{ref_eqpair}, we followed the evolution  of the optical depth in
the non-thermal  Comptonizing corona, the seed  photon temperature and
the  hard  state  compactness   parameters.   Our  results  show  that
independent of the  duration and the maximum flux of  an outburst, the
spectral  parameters  follow a  well  defined  and reproducible  track
during the long-high  and medium-low type outbursts  that are analysed
here.  A typical example of a  short-low type outburst is the May 1999
outburst.  This outburst was also  monitored with the RXTE/PCA and the
X-ray spectra was modeled with  the EQPAIR model by \citet{mac03spec}.
Comparing the spectral evolution during this outburst with the results
presented  here, we  see that  the optical  depth and  the compactness
parameters evolve in  a very similar way.  The only  difference in the
deduced spectral  parameters is a  slight increase in the  seed photon
temperature measured at the beginning  of the 1999 outburst \citep[see
  Figure 1.\ of][]{mac03spec}, which longer outbursts in our sample do
not show.

\begin{figure}
\centering
  \includegraphics[angle=0, scale=0.5]{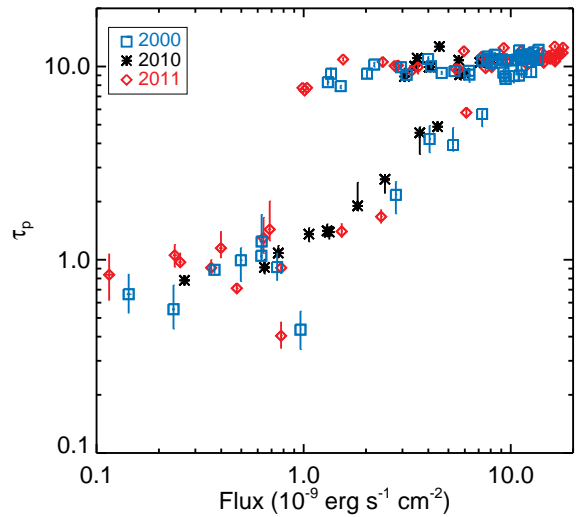}
     \caption{Evolution of  the scattering  optical
    depth of the corona as  a function of unabsorbed source flux. During the outbursts the parameters start 
    from the lower left corner, proceed to the upper right corner during the enhancement stage, 
    remain on the upper horizontal branch during the decay of the outburst and finally drop to the pre-outburst values.}
         \label{optev}
\end{figure}

In Figure \ref{optev}, we show the evolution of the optical depth with
flux. Once  an outburst commences, together  and almost simultaneously
with  the   flux,  the   scattering  optical   depth  of   the  corona
increases.  During  the  increase   the  Spearman's  rank  correlation
coefficient  between the  two parameters  is $0.92$.  This correlation
persists till the  optical depth reaches a saturation  point, which is
between 9 to 12 and independent from the peak flux of the outburst. 
The optical depth remains at the saturation values even during the 
decline of the outburst and drops to the pre-outburst values with a substantial time delay.

\begin{figure}
\centering
  \includegraphics[angle=0, scale=0.5]{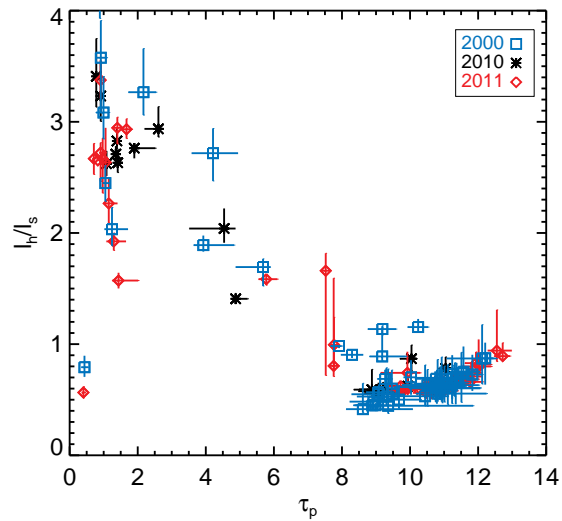}
     \caption{The relation between the compactness parameter $l_{\rm h}/l_{\rm s}$ and the optical depth
     of the corona. We have eliminated data with very large errorbars ($\Delta (l_{\rm h}/l_{\rm s})>1$).}
         \label{tau-l_h}
\end{figure}

We have also investigated the  relation between the compactness parameter
$l_{\rm h}/l_{\rm s}$ and the optical  depth of the corona as shown in
Figure~\ref{tau-l_h}. The bottom right corner of the figure corresponds to the HS state and upper left corner stands for the LH state.
For  $1<\tau<10$  the relation  is inverse and linear. This less populated region corresponds to the transitions between these two spectral 
states. As seen in Figure~\ref{plot_all}, $l_{\rm h}/l_{\rm s}$ shows similar time delay as $\tau$ does, 
which is reflected in the linear relation seen in Figure~\ref{tau-l_h}

\begin{figure}
\centering
  \includegraphics[angle=0, scale=0.5]{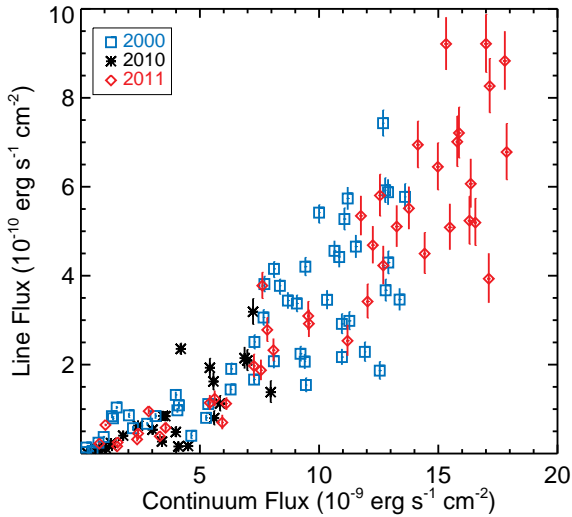}
  \caption{Variation of unabsorbed flux of the Gaussian iron line component as a
    with the EQPAIR model flux.}
         \label{flgausev}
\end{figure}

The time  evolution of the flux  of the Gaussian line  at $\approx6.5$
keV follows a well defined and  reproducible trend similar to the flux
obtained from the EQPAIR model.   In Figure \ref{flgausev} we show the
line flux  as a  function of  the continuum  flux.   The relation
  between the flux of the Fe line and the continuum flux does not seem
  to       be      exactly       linear       at      low       fluxes
  ($<6\times10^{-9}$\,erg~cm$^{-2}$~s$^{-1}$). As  the flux increases,
  however, the correlation becomes more linear.

Apart   from  some   exceptions  our   X-ray  spectral   analysis  and
observations of  Type-I X-ray bursts  indicate that the peak  flux and
duration of a  given outburst is relatively independent  of the region
in the  X-ray binary that determines  the shape of the  X-ray spectra.
The only difference between the  various classes of outbursts seems to
be the emitting radius of the  seed photons which determines the total
observed X-ray flux.

\section{Conclusion}
\label{conclude}

We have  presented a  broad classification of  the outbursts  of \aqls
based on the  duration and maximum flux. We identified  three types of
outbursts,  \emph{long-high}, \emph{medium-low}  and \emph{short-low}.
In  terms of  total  energy released,  the \emph{long-high}  outbursts
overwhelms the others.

We analysed the spectral evolution throughout three outbursts of \aqls
observed by RXTE, the 2000, 2010 and 2011, using the EQPAIR model. The
2010 outburst is  a member of the medium-low type  while 2000 and 2011
outbursts are members of the long-high type. These outbursts, together
with  the 1999  outburst analyzed  by \citet{mac03spec}  which is  of the
short-low  type, give  us an  opportunity  to discuss  three types  of
outbursts analysed with the same method.  Our spectral analysis showed
that all of the inferred parameters  evolve in a very similar way
regardless  of  the different  classes  of  the outbursts,  with  only
difference being the maximum flux. This implies that the ingredient in the
X-ray binary that  shapes the observed X-ray spectra is  not the cause
of the different outburst types.

Irradiation  of the  disc  modulates  the disc  size  involved in  the
outbursts  thereby  leading to  different  luminosities in different classes.  
We showed  a
correlation between the maximum flux and  the outer radius of the disc
inferred  by  \citet{cam+13}  which  demonstrates   the  role  of
irradiation as the main cause  for the different peak luminosities. We
argued that the transition to the  propeller or partial  accretion regime
induced by transition  to a geometrically thick  disc might contribute
to the diversity in the observed lightcurve morphology.


\section*{Acknowledgements}

This work is supported by The Scientific and Technological Council of 
TURKEY (TUBITAK) with the project number 112T105. We thank M.\ Ali Alpar and Ersin G\"{o}\u{g}\"{u}\c{s}
for their valuable comments and careful reading of the manuscript. TG acknowledges support from Bilim Akademisi - The Science Academy, Turkey 
under the BAGEP program. KYE thanks Ralph Neuh\"auser and Klaus Werner for hospitality during his visit to Jena and T\"ubingen, respectively.

\newpage

\begin{table*}
\center\tiny
\caption{Best fit parameters of EQPAIR model and the Gaussian line for
  2000 outburst.}
\begin{tabular}{cccccccccccccccccccccccc}
\hline
JD$-51000$	&	 $l_{\rm h} / l_{\rm s}$					&	 $kT_{\rm bb}$					&	 $\tau_{\rm p}$ &	$\chi^2$/d.o.f. &	 Line Energy & Line Width &	Flux$^{a}_{eqpair}$ & Flux$^{b}_{gauss}$ \\
(days) & & (eV) & & & (keV) & (keV) & &  \\ \hline
811.2856	&$	4.80	^{+	1.27	}_{-	1.15	}$ & $	658	^{+	292	}_{-	61	}$ & $	0.55	^{+	0.18	}_{-	0.12	}$ &	0.65	&$	5.80	^{+	0.33	}_{-	5.80	}$ & $	1.71	^{c		}_{-	0.36	}$ & $	0.235	^{+	0.004	}_{-	0.004	}$ & $	0.136	^{+	0.008	}_{-	0.008	}$\\
812.3495	&$	3.08	^{+	0.32	}_{-	0.24	}$ & $	719	^{+	62	}_{-	80	}$ & $	0.99	^{+	0.16	}_{-	0.22	}$ &	0.68	&$	6.34	^{+	0.19	}_{-	0.21	}$ & $	0.57	^{+	0.33	}_{-	0.29	}$ & $	0.497	^{+	0.004	}_{-	0.004	}$ & $	0.061	^{+	0.008	}_{-	0.008	}$\\
813.2773	&$	3.58	^{+	0.33	}_{-	0.49	}$ & $	714	^{+	101	}_{-	55	}$ & $	0.92	^{+	0.17	}_{-	0.14	}$ &	0.82	&$	5.80	^{+	0.24	}_{-	5.80	}$ & $	1.40	^{+	0.18	}_{-	0.27	}$ & $	0.745	^{+	0.004	}_{-	0.004	}$ & $	0.245	^{+	0.013	}_{-	0.015	}$\\
816.4667	&$	3.27	^{+	0.39	}_{-	0.21	}$ & $	684	^{+	220	}_{-	192	}$ & $	2.17	^{+	0.37	}_{-	0.44	}$ &	0.68	&$	6.07	^{+	0.17	}_{-	0.19	}$ & $	1.20	^{+	0.20	}_{-	0.18	}$ & $	2.787	^{+	0.008	}_{-	0.007	}$ & $	0.670	^{+	0.038	}_{-	0.037	}$\\
817.7979	&$	2.72	^{+	0.22	}_{-	0.25	}$ & $	624	^{+	161	}_{-	209	}$ & $	4.21	^{+	0.74	}_{-	0.63	}$ &	0.67	&$	6.04	^{+	0.17	}_{-	0.19	}$ & $	1.20	^{+	0.19	}_{-	0.17	}$ & $	4.055	^{+	0.011	}_{-	0.011	}$ & $	0.973	^{+	0.055	}_{-	0.054	}$\\
818.7868	&$	1.89	^{+	0.09	}_{-	0.06	}$ & $	856	^{+	39	}_{-	47	}$ & $	3.93	^{+	0.90	}_{-	0.28	}$ &	0.54	&$	6.52	^{+	0.14	}_{-	0.16	}$ & $	0.85	^{+	0.20	}_{-	0.18	}$ & $	5.260	^{+	0.011	}_{-	0.012	}$ & $	0.804	^{+	0.061	}_{-	0.059	}$\\
820.5169	&$	1.69	^{+	0.07	}_{-	0.17	}$ & $	795	^{+	110	}_{-	41	}$ & $	5.68	^{+	0.23	}_{-	0.81	}$ &	0.54	&$	6.38	^{+	0.14	}_{-	0.16	}$ & $	1.13	^{+	0.17	}_{-	0.15	}$ & $	7.273	^{+	0.009	}_{-	0.026	}$ & $	1.663	^{+	0.091	}_{-	0.089	}$\\
822.7745	&$	0.48	^{+	0.07	}_{-	0.08	}$ & $	715	^{+	84	}_{-	70	}$ & $	8.92	^{+	0.46	}_{-	0.70	}$ &	0.56	&$	6.29	^{+	0.24	}_{-	0.30	}$ & $	1.34	^{+	0.24	}_{-	0.21	}$ & $	10.334	^{+	0.026	}_{-	0.026	}$ & $	3.457	^{+	0.218	}_{-	0.211	}$\\
823.7689	&$	0.45	^{+	0.33	}_{-	0.07	}$ & $	820	^{+	75	}_{-	295	}$ & $	9.36	^{+	2.49	}_{-	0.63	}$ &	0.69	&$	6.72	^{+	0.20	}_{-	0.24	}$ & $	0.82	^{+	0.30	}_{-	0.25	}$ & $	12.544	^{+	0.032	}_{-	0.033	}$ & $	1.865	^{+	0.203	}_{-	0.193	}$\\
824.7611	&$	0.75	^{+	0.23	}_{-	0.13	}$ & $	592	^{+	104	}_{-	260	}$ & $	11.57	^{+	0.49	}_{-	0.40	}$ &	0.32	&$	6.26	^{+	0.22	}_{-	0.26	}$ & $	0.91	^{+	0.26	}_{-	0.23	}$ & $	11.919	^{+	0.038	}_{-	0.035	}$ & $	2.287	^{+	0.223	}_{-	0.213	}$\\
825.7550	&$	0.69	^{+	0.03	}_{-	0.11	}$ & $	547	^{+	221	}_{-	70	}$ & $	10.76	^{+	0.26	}_{-	0.62	}$ &	0.64	&$	5.81	^{+	0.21	}_{-	5.81	}$ & $	1.42	^{+	0.08	}_{-	0.17	}$ & $	10.000	^{+	0.029	}_{-	0.015	}$ & $	5.418	^{+	0.175	}_{-	0.235	}$\\
826.5129	&$	0.65	^{+	0.18	}_{-	0.13	}$ & $	653	^{+	107	}_{-	125	}$ & $	11.32	^{+	0.39	}_{-	0.26	}$ &	0.39	&$	6.25	^{+	0.18	}_{-	0.21	}$ & $	1.15	^{+	0.18	}_{-	0.17	}$ & $	12.902	^{+	0.034	}_{-	0.034	}$ & $	4.296	^{+	0.260	}_{-	0.252	}$\\
828.5344	&$	0.45	^{+	0.01	}_{-	0.05	}$ & $	755	^{+	54	}_{-	13	}$ & $	8.98	^{+	0.20	}_{-	0.59	}$ &	0.68	&$	6.61	^{+	0.18	}_{-	0.22	}$ & $	0.93	^{+	0.22	}_{-	0.20	}$ & $	10.958	^{+	0.030	}_{-	0.029	}$ & $	2.171	^{+	0.184	}_{-	0.175	}$\\
831.2024	&$	0.69	^{+	0.14	}_{-	0.11	}$ & $	615	^{+	88	}_{-	129	}$ & $	11.37	^{+	0.32	}_{-	0.19	}$ &	0.53	&$	5.96	^{+	0.20	}_{-	5.96	}$ & $	1.38	^{+	0.17	}_{-	0.16	}$ & $	11.061	^{+	0.028	}_{-	0.032	}$ & $	5.274	^{+	0.250	}_{-	0.244	}$\\
833.4586	&$	0.50	^{+	0.13	}_{-	0.08	}$ & $	697	^{+	58	}_{-	105	}$ & $	9.67	^{+	0.64	}_{-	0.89	}$ &	0.74	&$	6.52	^{+	0.17	}_{-	0.20	}$ & $	0.97	^{+	0.19	}_{-	0.17	}$ & $	11.268	^{+	0.033	}_{-	0.031	}$ & $	2.984	^{+	0.207	}_{-	0.198	}$\\
834.3003	&$	0.87	^{+	0.14	}_{-	0.24	}$ & $	523	^{+	160	}_{-	112	}$ & $	12.21	^{+	0.24	}_{-	0.25	}$ &	0.35	&$	5.92	^{+	0.19	}_{-	5.92	}$ & $	1.33	^{+	0.15	}_{-	0.16	}$ & $	13.604	^{+	0.036	}_{-	0.036	}$ & $	5.772	^{+	0.295	}_{-	0.289	}$\\
835.1903	&$	0.73	^{+	0.08	}_{-	0.19	}$ & $	529	^{+	154	}_{-	129	}$ & $	11.26	^{+	0.27	}_{-	0.19	}$ &	0.43	&$	5.88	^{+	0.18	}_{-	5.88	}$ & $	1.29	^{+	0.12	}_{-	0.15	}$ & $	7.705	^{+	0.022	}_{-	0.023	}$ & $	3.812	^{+	0.176	}_{-	0.172	}$\\
835.4761	&$	0.63	^{+	0.10	}_{-	0.17	}$ & $	651	^{+	145	}_{-	73	}$ & $	11.24	^{+	0.78	}_{-	1.01	}$ &	0.34	&$	6.06	^{+	0.19	}_{-	0.22	}$ & $	1.37	^{+	0.17	}_{-	0.15	}$ & $	12.771	^{+	0.034	}_{-	0.035	}$ & $	5.920	^{+	0.280	}_{-	0.274	}$\\
836.5150	&$	0.70	^{+	0.03	}_{-	0.24	}$ & $	596	^{+	203	}_{-	100	}$ & $	11.34	^{+	0.28	}_{-	0.37	}$ &	0.52	&$	5.88	^{+	0.19	}_{-	5.88	}$ & $	1.48	^{+	0.12	}_{-	0.15	}$ & $	12.683	^{+	0.035	}_{-	0.036	}$ & $	7.430	^{+	0.295	}_{-	0.290	}$\\
836.5906	&$	0.56	^{+	0.31	}_{-	0.07	}$ & $	734	^{+	61	}_{-	245	}$ & $	10.81	^{+	1.44	}_{-	0.64	}$ &	0.64	&$	6.47	^{+	0.19	}_{-	0.22	}$ & $	1.03	^{+	0.22	}_{-	0.19	}$ & $	13.376	^{+	0.036	}_{-	0.036	}$ & $	3.463	^{+	0.254	}_{-	0.244	}$\\
836.6590	&$	0.61	^{+	0.21	}_{-	0.15	}$ & $	647	^{+	133	}_{-	153	}$ & $	11.11	^{+	0.96	}_{-	1.03	}$ &	0.42	&$	6.05	^{+	0.21	}_{-	0.24	}$ & $	1.25	^{+	0.20	}_{-	0.18	}$ & $	11.543	^{+	0.034	}_{-	0.033	}$ & $	4.655	^{+	0.261	}_{-	0.255	}$\\
837.3107	&$	0.73	^{+	0.18	}_{-	0.15	}$ & $	605	^{+	111	}_{-	161	}$ & $	11.83	^{+	0.29	}_{-	0.42	}$ &	0.62	&$	5.95	^{+	0.23	}_{-	5.95	}$ & $	1.46	^{+	0.17	}_{-	0.18	}$ & $	12.892	^{+	0.035	}_{-	0.034	}$ & $	5.877	^{+	0.291	}_{-	0.285	}$\\
838.5825	&$	0.61	^{+	0.15	}_{-	0.15	}$ & $	622	^{+	135	}_{-	127	}$ & $	10.79	^{+	0.69	}_{-	1.27	}$ &	0.78	&$	6.11	^{+	0.18	}_{-	0.21	}$ & $	1.19	^{+	0.18	}_{-	0.16	}$ & $	10.843	^{+	0.031	}_{-	0.031	}$ & $	4.419	^{+	0.238	}_{-	0.231	}$\\
839.1655	&$	0.60	^{+	0.20	}_{-	0.12	}$ & $	680	^{+	113	}_{-	143	}$ & $	11.01	^{+	1.05	}_{-	1.05	}$ &	0.44	&$	6.34	^{+	0.18	}_{-	0.21	}$ & $	1.05	^{+	0.20	}_{-	0.18	}$ & $	12.792	^{+	0.034	}_{-	0.033	}$ & $	3.672	^{+	0.251	}_{-	0.242	}$\\
841.4233	&$	0.53	^{+	0.26	}_{-	0.08	}$ & $	661	^{+	84	}_{-	197	}$ & $	9.31	^{+	0.36	}_{-	0.87	}$ &	0.65	&$	6.48	^{+	0.18	}_{-	0.22	}$ & $	0.97	^{+	0.21	}_{-	0.19	}$ & $	9.220	^{+	0.027	}_{-	0.026	}$ & $	2.244	^{+	0.169	}_{-	0.162	}$\\
843.4235	&$	0.70	^{+	0.20	}_{-	0.19	}$ & $	597	^{+	154	}_{-	90	}$ & $	11.24	^{+	0.66	}_{-	1.08	}$ &	0.50	&$	5.93	^{+	0.21	}_{-	5.93	}$ & $	1.45	^{+	0.15	}_{-	0.17	}$ & $	11.201	^{+	0.030	}_{-	0.034	}$ & $	5.737	^{+	0.258	}_{-	0.253	}$\\
844.1205	&$	0.62	^{+	0.16	}_{-	0.12	}$ & $	646	^{+	103	}_{-	124	}$ & $	10.85	^{+	0.77	}_{-	0.34	}$ &	0.44	&$	6.13	^{+	0.19	}_{-	0.22	}$ & $	1.32	^{+	0.17	}_{-	0.16	}$ & $	10.645	^{+	0.031	}_{-	0.027	}$ & $	4.558	^{+	0.229	}_{-	0.223	}$\\
845.1457	&$	0.87	^{+	0.30	}_{-	0.23	}$ & $	516	^{+	162	}_{-	516	}$ & $	12.12	^{+	0.42	}_{-	1.04	}$ &	0.25	&$	6.06	^{+	0.24	}_{-	6.06	}$ & $	1.12	^{+	0.23	}_{-	0.21	}$ & $	10.955	^{+	0.033	}_{-	0.029	}$ & $	2.916	^{+	0.227	}_{-	0.219	}$\\
846.2087	&$	0.63	^{+	0.12	}_{-	0.15	}$ & $	606	^{+	134	}_{-	63	}$ & $	10.75	^{+	0.33	}_{-	0.99	}$ &	0.45	&$	6.08	^{+	0.19	}_{-	0.22	}$ & $	1.28	^{+	0.17	}_{-	0.16	}$ & $	9.423	^{+	0.027	}_{-	0.030	}$ & $	4.203	^{+	0.208	}_{-	0.203	}$\\
849.3941	&$	0.63	^{+	0.14	}_{-	0.19	}$ & $	573	^{+	177	}_{-	109	}$ & $	10.51	^{+	0.40	}_{-	0.70	}$ &	0.48	&$	5.82	^{+	0.25	}_{-	5.82	}$ & $	1.45	^{+	0.11	}_{-	0.20	}$ & $	8.108	^{+	0.025	}_{-	0.025	}$ & $	4.153	^{+	0.163	}_{-	0.197	}$\\
849.4632	&$	0.42	^{+	0.23	}_{-	0.03	}$ & $	810	^{+	38	}_{-	44	}$ & $	8.61	^{+	1.47	}_{-	0.49	}$ &	0.77	&$	6.71	^{+	0.21	}_{-	0.26	}$ & $	0.90	^{+	0.29	}_{-	0.24	}$ & $	9.447	^{+	0.029	}_{-	0.021	}$ & $	1.542	^{+	0.160	}_{-	0.151	}$\\
849.5347	&$	0.45	^{+	0.07	}_{-	0.04	}$ & $	774	^{+	49	}_{-	88	}$ & $	8.89	^{+	0.56	}_{-	0.55	}$ &	0.83	&$	6.60	^{+	0.21	}_{-	0.26	}$ & $	1.03	^{+	0.25	}_{-	0.21	}$ & $	9.398	^{+	0.023	}_{-	0.025	}$ & $	2.068	^{+	0.172	}_{-	0.164	}$\\
850.1023	&$	0.61	^{+	0.19	}_{-	0.11	}$ & $	646	^{+	95	}_{-	101	}$ & $	10.89	^{+	0.70	}_{-	0.70	}$ &	0.41	&$	6.20	^{+	0.19	}_{-	0.22	}$ & $	1.21	^{+	0.18	}_{-	0.16	}$ & $	9.067	^{+	0.025	}_{-	0.025	}$ & $	3.374	^{+	0.190	}_{-	0.185	}$\\
850.8640	&$	0.74	^{+	0.19	}_{-	0.26	}$ & $	544	^{+	153	}_{-	165	}$ & $	11.51	^{+	0.33	}_{-	0.26	}$ &	0.72	&$	5.81	^{+	0.24	}_{-	5.81	}$ & $	1.44	^{+	0.10	}_{-	0.18	}$ & $	8.355	^{+	0.025	}_{-	0.023	}$ & $	3.772	^{+	0.151	}_{-	0.190	}$\\
851.2877	&$	0.64	^{+	0.24	}_{-	0.16	}$ & $	617	^{+	151	}_{-	122	}$ & $	10.93	^{+	0.66	}_{-	0.81	}$ &	0.60	&$	6.09	^{+	0.21	}_{-	0.25	}$ & $	1.32	^{+	0.19	}_{-	0.17	}$ & $	8.692	^{+	0.025	}_{-	0.023	}$ & $	3.440	^{+	0.188	}_{-	0.183	}$\\
852.5877	&$	0.63	^{+	0.09	}_{-	0.16	}$ & $	592	^{+	150	}_{-	70	}$ & $	10.96	^{+	0.70	}_{-	0.47	}$ &	1.10	&$	6.01	^{+	0.23	}_{-	6.01	}$ & $	1.30	^{+	0.21	}_{-	0.19	}$ & $	7.669	^{+	0.023	}_{-	0.023	}$ & $	3.059	^{+	0.176	}_{-	0.172	}$\\
852.6531	&$	0.53	^{+	0.28	}_{-	0.09	}$ & $	694	^{+	77	}_{-	191	}$ & $	10.44	^{+	0.70	}_{-	1.00	}$ &	0.72	&$	6.47	^{+	0.21	}_{-	0.27	}$ & $	1.02	^{+	0.26	}_{-	0.22	}$ & $	8.104	^{+	0.024	}_{-	0.024	}$ & $	2.077	^{+	0.167	}_{-	0.159	}$\\
853.5833	&$	0.63	^{+	0.13	}_{-	0.13	}$ & $	596	^{+	115	}_{-	91	}$ & $	10.74	^{+	0.44	}_{-	0.68	}$ &	0.54	&$	6.11	^{+	0.22	}_{-	0.26	}$ & $	1.18	^{+	0.22	}_{-	0.20	}$ & $	7.288	^{+	0.022	}_{-	0.022	}$ & $	2.510	^{+	0.161	}_{-	0.156	}$\\
854.2723	&$	0.55	^{+	0.09	}_{-	0.10	}$ & $	621	^{+	101	}_{-	91	}$ & $	9.09	^{+	0.38	}_{-	0.81	}$ &	0.43	&$	6.22	^{+	0.25	}_{-	0.31	}$ & $	1.28	^{+	0.24	}_{-	0.21	}$ & $	6.314	^{+	0.019	}_{-	0.017	}$ & $	1.904	^{+	0.132	}_{-	0.126	}$\\
855.3036	&$	0.69	^{+	0.08	}_{-	0.07	}$ & $	454	^{+	68	}_{-	92	}$ & $	9.27	^{+	0.22	}_{-	0.23	}$ &	0.51	&$	6.61	^{+	0.29	}_{-	0.42	}$ & $	0.68	^{+	0.42	}_{-	0.35	}$ & $	4.639	^{+	0.012	}_{-	0.011	}$ & $	0.400	^{+	0.073	}_{-	0.066	}$\\
856.1874	&$	0.56	^{+	0.02	}_{-	0.07	}$ & $	647	^{+	73	}_{-	19	}$ & $	9.50	^{+	0.07	}_{-	0.26	}$ &	0.49	&$	6.48	^{+	0.22	}_{-	0.28	}$ & $	1.08	^{+	0.26	}_{-	0.22	}$ & $	6.289	^{+	0.018	}_{-	0.017	}$ & $	1.441	^{+	0.118	}_{-	0.112	}$\\
856.9696	&$	0.60	^{+	0.17	}_{-	0.01	}$ & $	577	^{+	15	}_{-	134	}$ & $	9.47	^{+	0.18	}_{-	0.04	}$ &	0.52	&$	6.40	^{+	0.23	}_{-	0.30	}$ & $	1.05	^{+	0.26	}_{-	0.23	}$ & $	5.350	^{+	0.016	}_{-	0.016	}$ & $	1.114	^{+	0.100	}_{-	0.095	}$\\
859.2409	&$	0.70	^{+	0.15	}_{-	0.16	}$ & $	492	^{+	82	}_{-	119	}$ & $	10.03	^{+	0.17	}_{-	0.22	}$ &	0.62	&$	6.26	^{+	0.23	}_{-	0.29	}$ & $	1.14	^{+	0.24	}_{-	0.21	}$ & $	4.122	^{+	0.013	}_{-	0.012	}$ & $	1.084	^{+	0.083	}_{-	0.079	}$\\
860.2152	&$	0.68	^{+	0.08	}_{-	0.11	}$ & $	527	^{+	94	}_{-	62	}$ & $	10.95	^{+	0.30	}_{-	0.44	}$ &	0.51	&$	5.80	^{+	0.32	}_{-	5.80	}$ & $	1.41	^{+	0.15	}_{-	0.27	}$ & $	3.989	^{+	0.012	}_{-	0.012	}$ & $	1.319	^{+	0.071	}_{-	0.087	}$\\
860.5474	&$	2.71	^{	c	}_{-	2.01	}$ & $	1	^{+	464	}_{-	1	}$ & $	9.92	^{+	0.37	}_{-	0.28	}$ &	0.73	&$	6.46	^{+	0.31	}_{-	0.42	}$ & $	1.10	^{+	0.37	}_{-	0.29	}$ & $	2.948	^{+	0.015	}_{-	0.019	}$ & $	0.494	^{+	0.108	}_{-	0.096	}$\\
861.1194	&$	0.57	^{+	0.03	}_{-	0.06	}$ & $	600	^{+	28	}_{-	171	}$ & $	9.06	^{+	0.18	}_{-	0.35	}$ &	0.67	&$	6.39	^{+	0.31	}_{-	0.45	}$ & $	1.39	^{+	0.36	}_{-	0.29	}$ & $	3.139	^{+	0.013	}_{-	0.008	}$ & $	0.844	^{+	0.065	}_{-	0.062	}$\\
861.8861	&$	1.15	^{+	0.07	}_{-	0.04	}$ & $	190	^{+	140	}_{-	26	}$ & $	10.24	^{+	0.29	}_{-	0.31	}$ &	0.72	&$	6.20	^{	c	}_{		}$ & $	1.49	^{+	0.17	}_{-	0.17	}$ & $	2.192	^{+	0.007	}_{-	0.008	}$ & $	0.570	^{+	0.048	}_{-	0.045	}$\\
862.0931	&$	1.14	^{+	0.03	}_{-	0.05	}$ & $	164	^{+	14	}_{-	39	}$ & $	9.19	^{+	0.40	}_{-	0.42	}$ &	0.82	&$	5.95	^{+	0.45	}_{-	0.73	}$ & $	1.73	^{+	0.47	}_{-	0.34	}$ & $	2.022	^{+	0.010	}_{-	0.005	}$ & $	0.860	^{+	0.050	}_{-	0.048	}$\\
863.2400	&$	0.98	^{+	0.01	}_{-	0.02	}$ & $	236	^{+	83	}_{-	236	}$ & $	7.91	^{+	0.08	}_{-	0.25	}$ &	0.67	&$	5.88	^{+	0.24	}_{-	5.88	}$ & $	1.69	^{	c	}_{-	0.17	}$ & $	1.509	^{+	0.004	}_{-	0.005	}$ & $	1.036	^{+	0.036	}_{-	0.035	}$\\
864.2082	&$	0.89	^{+	0.26	}_{-	0.04	}$ & $	303	^{+	165	}_{-	34	}$ & $	9.18	^{+	0.70	}_{-	0.37	}$ &	0.55	&$	5.84	^{+	0.44	}_{-	5.84	}$ & $	1.74	^{	c	}_{-	0.35	}$ & $	1.356	^{+	0.006	}_{-	0.006	}$ & $	0.793	^{+	0.037	}_{-	0.040	}$\\
864.2775	&$	0.90	^{+	0.04	}_{-	0.03	}$ & $	270	^{+	134	}_{-	194	}$ & $	8.30	^{+	0.33	}_{-	0.33	}$ &	0.64	&$	5.99	^{+	0.37	}_{-	5.99	}$ & $	1.64	^{	c	}_{-	0.31	}$ & $	1.308	^{+	0.006	}_{-	0.007	}$ & $	0.841	^{+	0.041	}_{-	0.040	}$\\
865.2722	&$	0.79	^{+	0.10	}_{-	0.09	}$ & $	320	^{+	42	}_{-	63	}$ & $	0.43	^{+	0.11	}_{-	0.09	}$ &	0.95	&$	5.88	^{+	0.21	}_{-	5.88	}$ & $	1.13	^{+	0.23	}_{-	0.24	}$ & $	0.965	^{+	0.005	}_{-	0.006	}$ & $	0.373	^{+	0.025	}_{-	0.024	}$\\
866.9225	&$	2.03	^{+	0.20	}_{-	0.15	}$ & $	760	^{+	107	}_{-	261	}$ & $	1.24	^{+	0.47	}_{-	0.22	}$ &	1.02	&$	6.08	^{+	0.28	}_{-	0.29	}$ & $	0.25	^{+	0.46	}_{-	0.00	}$ & $	0.627	^{+	0.013	}_{-	0.012	}$ & $	0.049	^{+	0.022	}_{-	0.019	}$\\
866.9779	&$	2.45	^{+	0.16	}_{-	0.16	}$ & $	827	^{+	42	}_{-	40	}$ & $	1.05	^{+	0.11	}_{-	0.07	}$ &	2.21	&$	5.60	^{	c	}_{		}$ & $	0.93	^{	c	}_{		}$ & $	0.625	^{+	0.006	}_{-	0.007	}$ & $	0.065	^{	c	}_{		}$\\
868.2361	&$	4.89	^{+	0.88	}_{-	0.96	}$ & $	1049	^{+	16	}_{-	107	}$ & $	0.89	^{+	0.04	}_{-	0.04	}$ &	0.72	&$	5.87	^{	c	}_{		}$ & $	0.25	^{	c	}_{		}$ & $	0.371	^{+	0.002	}_{-	0.007	}$ & $	0.000	^{+	0.011	}_{		}$\\
869.4744	&$	4.68	^{+	1.81	}_{-	0.95	}$ & $	714	^{+	191	}_{-	58	}$ & $	0.66	^{+	0.18	}_{-	0.14	}$ &	0.62	&$	6.50	^{+	0.27	}_{-	0.33	}$ & $	0.82	^{+	0.48	}_{-	0.34	}$ & $	0.143	^{+	0.003	}_{-	0.003	}$ & $	0.034	^{+	0.005	}_{-	0.005	}$\\
870.3233	&$	7.51	^{+	3.92	}_{-	2.80	}$ & $	1071	^{+	119	}_{-	205	}$ & $	1.11	^{+	0.16	}_{-	0.16	}$ &	1.50	&$	5.60	^{	c	}_{		}$ & $	1.75	^{	c	}_{		}$ & $	0.078	^{+	0.004	}_{-	0.004	}$ & $	0.100	^{	c	}_{		}$\\
871.2461	&$	12.86	^{+	13.06	}_{-	7.65	}$ & $	724	^{+	600	}_{-	246	}$ & $	0.00	^{+	1.41	}_{-	0.00	}$ &	0.54	&$	6.16	^{	c	}_{		}$ & $	1.37	^{	c	}_{		}$ & $	0.041	^{+	0.002	}_{-	0.003	}$ & $	0.012	^{+	0.007	}_{-	0.008	}$\\
872.1597	&$	3.16	^{	c	}_{		}$ & $	489	^{+	336	}_{-	140	}$ & $	0.00	^{	c	}_{		}$ &	0.78	&$	6.89	^{	c	}_{		}$ & $	0.94	^{	c	}_{		}$ & $	0.020	^{+	0.004	}_{-	0.003	}$ & $	0.011	^{+	0.008	}_{-	0.007	}$\\
874.3083	&$	62.85	^{+	70.60	}_{-	57.91	}$ & $	1056	^{+	469	}_{-	555	}$ & $	0.00	^{	c	}_{		}$ &	0.85	&$	6.39	^{	c	}_{		}$ & $	0.25	^{	c	}_{		}$ & $	0.015	^{+	0.002	}_{-	0.002	}$ & $	0.002	^{+	0.003	}_{	c	}$\\
876.3387	&$	19.23	^{	c	}_{		}$ & $	331	^{	c	}_{		}$ & $	7.15	^{	c	}_{		}$ &	0.74	&$	5.60	^{	c	}_{		}$ & $	0.25	^{	c	}_{		}$ & $	0.013	^{+	0.000	}_{-	0.000	}$ & $	0.003	^{+	0.004	}_{	c	}$\\
878.1575	&$	8.32	^{	c	}_{		}$ & $	746	^{+	528	}_{-	286	}$ & $	0.00	^{	c	}_{		}$ &	1.20	&$	5.61	^{	c	}_{		}$ & $	1.00	^{	c	}_{		}$ & $	0.011	^{+	0.003	}_{-	0.002	}$ & $	0.000	^{	c	}_{		}$\\ \hline
\end{tabular}
\footnotesize{ 
\begin{flushleft} $^{a}$ Unabsorbed flux of the EQPAIR model in units of $10^{-9}~{\rm erg~ s^{-1}~cm^{-2}}$. \\
$^b$ Unabsorbed flux of the Gaussian component in units of $10^{-12}~{\rm erg~s^{-1}~cm^{-2}}$.\\
$^c$ Errors bars could not be constrained either due to low flux levels or degeneracies between model parameters. 
\end{flushleft} }
\label{tab:par.2000}
\end{table*}

\pagebreak
\newpage

\begin{table*}
\caption{Same as Table \ref{tab:par.2000} but for 2010 outburst.}
\begin{center}
\tiny
\begin{tabular}{ccccccccccccccccccccccccccc}
\hline
JD$-55000$	&	 $l_{\rm h} / l_{\rm s}$					&	 $kT_{\rm bb}$					&	 $\tau_{\rm p}$ &	$\chi^2$/d.o.f. &	 Line Energy & Line Width &	Flux$^{a}_{eqpair}$ & Flux$^{b}_{gauss}$ \\
(days) & & (eV) & & & (keV) & (keV) &  &  \\ \hline
426.2689	&$	2.71	^{+	0.07	}_{-	0.06	}$ & $	784	^{+	53	}_{-	50	}$ & $	1.36	^{+	0.12	}_{-	0.11	}$ &	2.12	&$	5.60	^{+	0.11	}_{-	5.60	}$ & $	1.01	^{+	0.22	}_{-	0.21	}$ & $	1.048	^{+	0.006	}_{-	0.007	}$ & $	0.127	^{+	0.025	}_{-	0.024	}$\\
426.9906	&$	2.63	^{+	0.12	}_{-	0.09	}$ & $	740	^{+	57	}_{-	58	}$ & $	1.41	^{+	0.10	}_{-	0.12	}$ &	1.14	&$	5.60	^{+	0.10	}_{-	5.60	}$ & $	1.24	^{+	0.20	}_{-	0.19	}$ & $	1.275	^{+	0.009	}_{-	0.008	}$ & $	0.235	^{+	0.031	}_{-	0.031	}$\\
428.8847	&$	2.76	^{+	0.04	}_{-	0.09	}$ & $	672	^{+	47	}_{-	77	}$ & $	1.90	^{+	0.11	}_{-	0.62	}$ &	1.92	&$	5.60	^{+	0.09	}_{-	5.60	}$ & $	1.23	^{+	0.14	}_{-	0.14	}$ & $	1.782	^{+	0.006	}_{-	0.006	}$ & $	0.403	^{+	0.022	}_{-	0.031	}$\\
430.8469	&$	2.83	^{+	0.11	}_{-	0.07	}$ & $	795	^{+	66	}_{-	14	}$ & $	1.39	^{+	0.12	}_{-	0.10	}$ &	2.06	&$	5.60	^{+	0.07	}_{-	5.60	}$ & $	0.97	^{+	0.20	}_{-	0.19	}$ & $	1.313	^{+	0.004	}_{-	0.019	}$ & $	0.169	^{+	0.029	}_{-	0.029	}$\\
433.8503	&$	2.62	^{+	0.12	}_{-	0.11	}$ & $	718	^{+	50	}_{-	51	}$ & $	1.08	^{+	0.05	}_{-	0.10	}$ &	1.59	&$	5.60	^{+	0.13	}_{-	5.60	}$ & $	0.94	^{+	0.25	}_{-	0.25	}$ & $	0.746	^{+	0.006	}_{-	0.007	}$ & $	0.089	^{+	0.019	}_{-	0.020	}$\\
436.3373	&$	3.41	^{+	0.34	}_{-	0.28	}$ & $	755	^{+	30	}_{-	34	}$ & $	0.78	^{+	0.04	}_{-	0.04	}$ &	0.82	&$	6.68	^{+	0.25	}_{-	0.24	}$ & $	0.25	^{+	0.42	}_{-	0.25	}$ & $	0.264	^{+	0.004	}_{-	0.004	}$ & $	0.022	^{+	0.005	}_{-	0.005	}$\\
441.8428	&$	3.23	^{+	0.31	}_{-	0.23	}$ & $	756	^{+	51	}_{-	44	}$ & $	0.91	^{+	0.07	}_{-	0.09	}$ &	0.68	&$	6.02	^{+	0.27	}_{-	0.35	}$ & $	0.76	^{+	0.51	}_{-	0.45	}$ & $	0.636	^{+	0.010	}_{-	0.010	}$ & $	0.110	^{+	0.026	}_{-	0.025	}$\\
447.5829	&$	2.94	^{+	0.20	}_{-	0.07	}$ & $	582	^{+	93	}_{-	102	}$ & $	2.61	^{+	0.41	}_{-	0.11	}$ &	0.80	&$	5.60	^{+	0.16	}_{-	5.60	}$ & $	1.35	^{+	0.14	}_{-	0.17	}$ & $	2.405	^{+	0.006	}_{-	0.007	}$ & $	0.629	^{+	0.031	}_{-	0.032	}$\\
449.3527	&$	2.04	^{+	0.18	}_{-	0.13	}$ & $	619	^{+	84	}_{-	58	}$ & $	4.54	^{+	1.03	}_{-	0.34	}$ &	0.83	&$	6.21	^{+	0.17	}_{-	0.19	}$ & $	1.08	^{+	0.23	}_{-	0.20	}$ & $	3.552	^{+	0.019	}_{-	0.011	}$ & $	0.841	^{+	0.091	}_{-	0.088	}$\\
450.0526	&$	1.41	^{+	0.05	}_{-	0.05	}$ & $	742	^{+	33	}_{-	41	}$ & $	4.88	^{+	0.14	}_{-	0.35	}$ &	1.10	&$	5.88	^{+	0.17	}_{-	0.19	}$ & $	1.62	^{+	-1.62	}_{-	0.17	}$ & $	4.197	^{+	0.021	}_{-	0.020	}$ & $	2.355	^{+	0.128	}_{-	0.126	}$\\
451.6399	&$	0.56	^{+	0.09	}_{-	0.04	}$ & $	589	^{+	37	}_{-	80	}$ & $	9.07	^{+	0.41	}_{-	0.83	}$ &	0.50	&$	6.54	^{+	0.23	}_{-	0.29	}$ & $	0.78	^{+	0.30	}_{-	0.25	}$ & $	5.596	^{+	0.031	}_{-	0.032	}$ & $	0.796	^{+	0.167	}_{-	0.151	}$\\
452.6867	&$	0.64	^{+	0.07	}_{-	0.06	}$ & $	570	^{+	49	}_{-	57	}$ & $	10.74	^{+	0.45	}_{-	0.45	}$ &	0.72	&$	5.87	^{+	0.30	}_{-	5.87	}$ & $	1.49	^{+	0.25	}_{-	0.24	}$ & $	7.229	^{+	0.036	}_{-	0.035	}$ & $	3.189	^{+	0.294	}_{-	0.283	}$\\
453.1981	&$	0.59	^{+	0.05	}_{-	0.05	}$ & $	643	^{+	64	}_{-	44	}$ & $	10.85	^{+	0.53	}_{-	0.49	}$ &	0.58	&$	6.38	^{+	0.22	}_{-	0.28	}$ & $	0.82	^{+	0.29	}_{-	0.25	}$ & $	7.970	^{+	0.038	}_{-	0.038	}$ & $	1.381	^{+	0.256	}_{-	0.233	}$\\
454.3926	&$	0.58	^{+	0.06	}_{-	0.05	}$ & $	588	^{+	48	}_{-	54	}$ & $	9.23	^{+	0.74	}_{-	0.61	}$ &	0.79	&$	6.50	^{+	0.20	}_{-	0.26	}$ & $	0.87	^{+	0.28	}_{-	0.24	}$ & $	5.844	^{+	0.026	}_{-	0.032	}$ & $	1.125	^{+	0.181	}_{-	0.166	}$\\
455.1530	&$	0.62	^{+	0.06	}_{-	0.08	}$ & $	595	^{+	39	}_{-	48	}$ & $	10.64	^{+	0.56	}_{-	0.42	}$ &	0.73	&$	6.27	^{+	0.20	}_{-	0.25	}$ & $	1.08	^{+	0.23	}_{-	0.20	}$ & $	6.988	^{+	0.033	}_{-	0.033	}$ & $	2.094	^{+	0.243	}_{-	0.228	}$\\
456.2053	&$	0.60	^{+	0.06	}_{-	0.05	}$ & $	606	^{+	39	}_{-	52	}$ & $	10.47	^{+	0.49	}_{-	0.51	}$ &	0.53	&$	6.30	^{+	0.24	}_{-	0.31	}$ & $	1.19	^{+	0.26	}_{-	0.22	}$ & $	6.873	^{+	0.033	}_{-	0.034	}$ & $	2.147	^{+	0.252	}_{-	0.237	}$\\
457.5870	&$	0.59	^{+	0.06	}_{-	0.08	}$ & $	567	^{+	80	}_{-	56	}$ & $	9.20	^{+	0.70	}_{-	0.53	}$ &	0.66	&$	6.22	^{+	0.28	}_{-	0.37	}$ & $	1.25	^{+	0.29	}_{-	0.25	}$ & $	5.568	^{+	0.028	}_{-	0.028	}$ & $	1.621	^{+	0.204	}_{-	0.191	}$\\
460.9298	&$	2.40	^{+	0.04	}_{-	1.59	}$ & $	4	^{+	411	}_{-	4	}$ & $	9.93	^{+	0.20	}_{-	0.26	}$ &	0.46	&$	6.38	^{+	0.55	}_{-	6.38	}$ & $	0.32	^{+	1.14	}_{-	0.32	}$ & $	4.096	^{+	0.018	}_{-	0.020	}$ & $	0.160	^{+	0.075	}_{-	0.056	}$\\
463.5337	&$	0.64	^{+	0.07	}_{-	0.06	}$ & $	571	^{+	51	}_{-	29	}$ & $	10.78	^{+	0.49	}_{-	0.48	}$ &	0.76	&$	5.76	^{+	0.41	}_{-	5.76	}$ & $	1.50	^{+	0.22	}_{-	0.34	}$ & $	5.418	^{+	0.027	}_{-	0.027	}$ & $	1.930	^{+	0.212	}_{-	0.220	}$\\
466.0262	&$	0.62	^{+	0.07	}_{-	0.06	}$ & $	560	^{+	49	}_{-	55	}$ & $	9.94	^{+	0.55	}_{-	0.53	}$ &	0.76	&$	6.56	^{+	0.23	}_{-	0.32	}$ & $	0.69	^{+	0.41	}_{-	0.33	}$ & $	3.987	^{+	0.016	}_{-	0.022	}$ & $	0.485	^{+	0.123	}_{-	0.108	}$\\
467.0082	&$	0.95	^{+	5.14	}_{-	0.14	}$ & $	407	^{+	81	}_{-	407	}$ & $	12.68	^{+	0.32	}_{-	0.29	}$ &	0.59	&$	6.38	^{+	0.36	}_{-	0.67	}$ & $	0.25	^{+	0.80	}_{-	0.25	}$ & $	4.505	^{+	0.014	}_{-	0.014	}$ & $	0.169	^{+	0.065	}_{-	0.050	}$\\
468.9646	&$	0.59	^{+	0.18	}_{-	0.05	}$ & $	528	^{+	50	}_{-	152	}$ & $	8.88	^{+	0.53	}_{-	0.93	}$ &	0.93	&$	7.00	^{+	-7.00	}_{-	0.36	}$ & $	1.75	^{+	-1.75	}_{-	0.20	}$ & $	3.015	^{+	0.022	}_{-	0.018	}$ & $	0.537	^{+	0.083	}_{-	0.078	}$\\
472.7545	&$	0.87	^{+	0.12	}_{-	0.11	}$ & $	401	^{+	80	}_{-	77	}$ & $	10.04	^{+	0.34	}_{-	0.15	}$ &	0.95	&$	6.79	^{+	-6.79	}_{-	0.24	}$ & $	0.45	^{+	0.37	}_{-	0.45	}$ & $	3.408	^{+	0.010	}_{-	0.011	}$ & $	0.270	^{+	0.047	}_{-	0.044	}$\\
474.4141	&$	0.78	^{+	0.10	}_{-	0.07	}$ & $	440	^{+	50	}_{-	80	}$ & $	11.05	^{+	0.21	}_{-	0.19	}$ &	0.98	&$	6.06	^{+	0.36	}_{-	6.06	}$ & $	1.33	^{+	0.36	}_{-	0.31	}$ & $	3.452	^{+	0.005	}_{-	0.010	}$ & $	0.847	^{+	0.076	}_{-	0.073	}$\\ \hline
\end{tabular}
\end{center}
\footnotesize{ 
\begin{flushleft} $^{a}$ Unabsorbed flux of the EQPAIR model in units of $10^{-9}~{\rm erg~s^{-1}~cm^{-2}}$. \\
$^b$ Unabsorbed flux of the Gaussian component in units of $10^{-12}~{\rm erg~s^{-1}~cm^{-2}}$.\\
$^c$ Errors bars could not be constrained either due to low flux levels or degeneracies between model parameters. 
\end{flushleft} }

\label{tab:par.2010}
\end{table*}

\pagebreak
\newpage

\begin{table*}
\tiny
\caption{Same as Table \ref{tab:par.2000} but for 2011 outburst.}
\begin{center}
\begin{tabular}{ccccccccccccccccccccccccccc}
\hline
JD$-55000$	&	 $l_{\rm h} / l_{\rm s}$					&	 $kT_{\rm bb}$					&	 $\tau_{\rm p}$ &	$\chi^2$/d.o.f. &	 Line Energy &
Line Width &	Flux$^{a}_{eqpair}$ & Flux$^{b}_{gauss}$ \\
(days) & & (eV) & & & (keV) & (keV) & &  \\ \hline
849.3771	&$	3.38	^{+	0.18	}_{-	0.16	}$ & $	706	^{+	32	}_{-	34	}$ & $	0.91	^{+	0.06	}_{-	0.06	}$ &	0.79	&$	5.81	^{+	0.30	}_{-	5.81	}$ & $	1.46	^{	c	}_{	0.30	}$ & $	0.779	^{+	0.007	}_{-	0.007	}$ & $	0.225	^{+	0.026	}_{-	0.025	}$\\
851.0042	&$	2.93	^{+	0.09	}_{-	0.08	}$ & $	816	^{+	47	}_{-	39	}$ & $	1.67	^{+	0.15	}_{-	0.13	}$ &	0.62	&$	6.38	^{+	0.17	}_{-	0.20	}$ & $	0.82	^{+	0.25	}_{-	0.23	}$ & $	2.366	^{+	0.013	}_{-	0.011	}$ & $	0.320	^{+	0.049	}_{-	0.046	}$\\
852.4341	&$	2.95	^{+	0.10	}_{-	0.14	}$ & $	754	^{+	47	}_{-	65	}$ & $	1.40	^{+	0.15	}_{-	0.10	}$ &	0.56	&$	6.40	^{+	0.20	}_{-	0.22	}$ & $	0.62	^{+	0.29	}_{-	0.26	}$ & $	1.527	^{+	0.010	}_{-	0.011	}$ & $	0.164	^{+	0.036	}_{-	0.034	}$\\
855.5781	&$	0.61	^{+	0.05	}_{-	0.04	}$ & $	628	^{+	38	}_{-	39	}$ & $	10.89	^{+	0.33	}_{-	0.35	}$ &	0.66	&$	6.04	^{+	0.19	}_{-	0.23	}$ & $	1.21	^{+	0.18	}_{-	0.17	}$ & $	13.256	^{+	0.059	}_{-	0.058	}$ & $	5.105	^{+	0.477	}_{-	0.457	}$\\
858.1831	&$	0.60	^{+	0.04	}_{-	0.04	}$ & $	608	^{+	33	}_{-	38	}$ & $	10.22	^{+	0.33	}_{-	0.31	}$ &	0.43	&$	6.24	^{+	0.18	}_{-	0.21	}$ & $	1.00	^{+	0.19	}_{-	0.17	}$ & $	12.033	^{+	0.054	}_{-	0.054	}$ & $	3.419	^{+	0.394	}_{-	0.371	}$\\
859.6397	&$	1.58	^{+	0.04	}_{-	0.05	}$ & $	835	^{+	22	}_{-	24	}$ & $	5.78	^{+	0.32	}_{-	0.22	}$ &	0.67	&$	6.22	^{+	0.18	}_{-	0.20	}$ & $	1.08	^{+	0.22	}_{-	0.19	}$ & $	6.099	^{+	0.023	}_{-	0.024	}$ & $	1.123	^{+	0.131	}_{-	0.126	}$\\
861.2436	&$	0.83	^{+	0.10	}_{-	0.07	}$ & $	586	^{+	49	}_{-	64	}$ & $	11.94	^{+	0.37	}_{-	0.32	}$ &	0.67	&$	6.12	^{+	0.26	}_{-	0.30	}$ & $	1.09	^{+	0.26	}_{-	0.23	}$ & $	17.118	^{+	0.071	}_{-	0.071	}$ & $	3.932	^{+	0.574	}_{-	0.542	}$\\
862.5005	&$	0.66	^{+	0.06	}_{-	0.05	}$ & $	617	^{+	41	}_{-	48	}$ & $	11.56	^{+	0.35	}_{-	0.37	}$ &	0.64	&$	5.97	^{+	0.18	}_{-	0.20	}$ & $	1.26	^{+	0.16	}_{-	0.15	}$ & $	15.866	^{+	0.070	}_{-	0.070	}$ & $	7.209	^{+	0.582	}_{-	0.562	}$\\
863.3370	&$	0.68	^{+	0.06	}_{-	0.05	}$ & $	624	^{+	43	}_{-	49	}$ & $	11.67	^{+	0.35	}_{-	0.35	}$ &	0.44	&$	6.02	^{+	0.17	}_{-	0.19	}$ & $	1.30	^{+	0.15	}_{-	0.14	}$ & $	17.155	^{+	0.074	}_{-	0.074	}$ & $	8.264	^{+	0.620	}_{-	0.600	}$\\
864.2519	&$	0.62	^{+	0.05	}_{-	0.05	}$ & $	626	^{+	41	}_{-	40	}$ & $	11.00	^{+	0.35	}_{-	0.39	}$ &	0.56	&$	6.09	^{+	0.17	}_{-	0.20	}$ & $	1.19	^{+	0.16	}_{-	0.15	}$ & $	14.981	^{+	0.068	}_{-	0.067	}$ & $	6.448	^{+	0.539	}_{-	0.519	}$\\
865.4324	&$	0.56	^{+	0.04	}_{-	0.04	}$ & $	691	^{+	36	}_{-	30	}$ & $	10.64	^{+	0.37	}_{-	0.42	}$ &	0.49	&$	6.21	^{+	0.18	}_{-	0.21	}$ & $	1.19	^{+	0.18	}_{-	0.16	}$ & $	16.361	^{+	0.071	}_{-	0.072	}$ & $	6.067	^{+	0.562	}_{-	0.536	}$\\
866.2953	&$	0.66	^{+	0.06	}_{-	0.05	}$ & $	655	^{+	41	}_{-	43	}$ & $	11.77	^{+	0.38	}_{-	0.41	}$ &	0.41	&$	5.88	^{+	0.20	}_{-	0.23	}$ & $	1.40	^{+	0.17	}_{-	0.16	}$ & $	17.792	^{+	0.077	}_{-	0.077	}$ & $	8.830	^{+	0.668	}_{-	0.649	}$\\
867.2469	&$	0.61	^{+	0.05	}_{-	0.05	}$ & $	635	^{+	42	}_{-	39	}$ & $	11.04	^{+	0.36	}_{-	0.42	}$ &	0.75	&$	6.07	^{+	0.18	}_{-	0.21	}$ & $	1.24	^{+	0.17	}_{-	0.16	}$ & $	15.794	^{+	0.071	}_{-	0.071	}$ & $	7.012	^{+	0.580	}_{-	0.558	}$\\
867.9782	&$	0.68	^{+	0.07	}_{-	0.06	}$ & $	620	^{+	42	}_{-	52	}$ & $	11.53	^{+	0.39	}_{-	0.44	}$ &	0.52	&$	5.81	^{+	0.20	}_{-	5.81	}$ & $	1.42	^{+	0.17	}_{-	0.16	}$ & $	17.001	^{+	0.077	}_{-	0.077	}$ & $	9.215	^{+	0.660	}_{-	0.643	}$\\
869.2827	&$	0.63	^{+	0.05	}_{-	0.05	}$ & $	616	^{+	42	}_{-	40	}$ & $	11.07	^{+	0.34	}_{-	0.39	}$ &	0.41	&$	6.08	^{+	0.18	}_{-	0.21	}$ & $	1.19	^{+	0.17	}_{-	0.15	}$ & $	13.766	^{+	0.065	}_{-	0.060	}$ & $	5.515	^{+	0.490	}_{-	0.470	}$\\
871.2931	&$	0.68	^{+	0.07	}_{-	0.06	}$ & $	592	^{+	47	}_{-	50	}$ & $	11.39	^{+	0.37	}_{-	0.40	}$ &	0.53	&$	5.77	^{+	0.18	}_{-	5.77	}$ & $	1.41	^{+	0.15	}_{-	0.14	}$ & $	15.326	^{+	0.070	}_{-	0.070	}$ & $	9.212	^{+	0.602	}_{-	0.586	}$\\
872.0839	&$	0.57	^{+	0.03	}_{-	0.04	}$ & $	712	^{+	39	}_{-	26	}$ & $	10.92	^{+	0.36	}_{-	0.48	}$ &	0.62	&$	6.35	^{+	0.18	}_{-	0.21	}$ & $	1.09	^{+	0.19	}_{-	0.17	}$ & $	16.554	^{+	0.070	}_{-	0.072	}$ & $	5.194	^{+	0.545	}_{-	0.515	}$\\
873.0034	&$	0.94	^{+	0.37	}_{-	0.10	}$ & $	516	^{+	61	}_{-	115	}$ & $	12.55	^{+	0.44	}_{-	0.27	}$ &	0.74	&$	5.79	^{+	0.22	}_{-	5.79	}$ & $	1.28	^{+	0.18	}_{-	0.17	}$ & $	17.867	^{+	0.079	}_{-	0.066	}$ & $	6.780	^{+	0.642	}_{-	0.623	}$\\
875.6732	&$	0.58	^{+	0.05	}_{-	0.04	}$ & $	684	^{+	32	}_{-	41	}$ & $	11.00	^{+	0.40	}_{-	0.49	}$ &	0.62	&$	6.31	^{+	0.20	}_{-	0.24	}$ & $	1.17	^{+	0.21	}_{-	0.19	}$ & $	15.484	^{+	0.068	}_{-	0.067	}$ & $	5.087	^{+	0.532	}_{-	0.504	}$\\
876.8402	&$	0.89	^{+	0.12	}_{-	0.03	}$ & $	548	^{+	36	}_{-	66	}$ & $	12.72	^{+	0.21	}_{-	0.25	}$ &	0.57	&$	6.13	^{+	0.21	}_{-	0.24	}$ & $	1.20	^{+	0.20	}_{-	0.18	}$ & $	16.308	^{+	0.067	}_{-	0.066	}$ & $	5.238	^{+	0.551	}_{-	0.528	}$\\
877.4295	&$	0.56	^{+	0.04	}_{-	0.03	}$ & $	689	^{+	30	}_{-	36	}$ & $	10.47	^{+	0.33	}_{-	0.38	}$ &	0.56	&$	6.33	^{+	0.18	}_{-	0.21	}$ & $	1.07	^{+	0.19	}_{-	0.17	}$ & $	14.428	^{+	0.063	}_{-	0.062	}$ & $	4.497	^{+	0.474	}_{-	0.448	}$\\
877.8880	&$	0.74	^{+	0.08	}_{-	0.06	}$ & $	572	^{+	49	}_{-	55	}$ & $	11.80	^{+	0.34	}_{-	0.35	}$ &	0.62	&$	5.82	^{+	0.19	}_{-	5.82	}$ & $	1.36	^{+	0.17	}_{-	0.15	}$ & $	14.142	^{+	0.063	}_{-	0.062	}$ & $	6.943	^{+	0.532	}_{-	0.517	}$\\
879.5833	&$	0.64	^{+	0.06	}_{-	0.05	}$ & $	601	^{+	41	}_{-	50	}$ & $	11.10	^{+	0.37	}_{-	0.38	}$ &	0.56	&$	5.86	^{+	0.22	}_{-	5.86	}$ & $	1.37	^{+	0.20	}_{-	0.18	}$ & $	11.754	^{+	0.054	}_{-	0.054	}$ & $	5.345	^{+	0.451	}_{-	0.436	}$\\
880.6295	&$	0.61	^{+	0.05	}_{-	0.05	}$ & $	631	^{+	40	}_{-	41	}$ & $	10.97	^{+	0.37	}_{-	0.40	}$ &	0.73	&$	6.19	^{+	0.18	}_{-	0.21	}$ & $	1.18	^{+	0.18	}_{-	0.16	}$ & $	12.253	^{+	0.055	}_{-	0.055	}$ & $	4.687	^{+	0.430	}_{-	0.410	}$\\
882.6413	&$	0.71	^{+	0.09	}_{-	0.06	}$ & $	580	^{+	49	}_{-	62	}$ & $	11.71	^{+	0.41	}_{-	0.42	}$ &	0.59	&$	5.89	^{+	0.20	}_{-	0.23	}$ & $	1.28	^{+	0.18	}_{-	0.17	}$ & $	12.552	^{+	0.057	}_{-	0.057	}$ & $	5.804	^{+	0.478	}_{-	0.462	}$\\
883.5596	&$	0.68	^{+	0.06	}_{-	0.06	}$ & $	607	^{+	41	}_{-	49	}$ & $	11.41	^{+	0.37	}_{-	0.38	}$ &	0.61	&$	6.14	^{+	0.20	}_{-	0.24	}$ & $	1.16	^{+	0.21	}_{-	0.19	}$ & $	12.691	^{+	0.054	}_{-	0.057	}$ & $	4.229	^{+	0.443	}_{-	0.422	}$\\
884.6704	&$	0.63	^{+	0.05	}_{-	0.05	}$ & $	591	^{+	43	}_{-	46	}$ & $	10.71	^{+	0.33	}_{-	0.37	}$ &	0.87	&$	6.20	^{+	0.19	}_{-	0.22	}$ & $	1.08	^{+	0.19	}_{-	0.17	}$ & $	9.544	^{+	0.043	}_{-	0.044	}$ & $	3.094	^{+	0.328	}_{-	0.311	}$\\
885.8371	&$	0.62	^{+	0.04	}_{-	0.04	}$ & $	607	^{+	32	}_{-	32	}$ & $	10.52	^{+	0.34	}_{-	0.27	}$ &	0.60	&$	6.28	^{+	0.18	}_{-	0.22	}$ & $	1.08	^{+	0.20	}_{-	0.18	}$ & $	9.571	^{+	0.041	}_{-	0.041	}$ & $	2.921	^{+	0.317	}_{-	0.299	}$\\
886.8228	&$	0.80	^{+	0.11	}_{-	0.07	}$ & $	554	^{+	48	}_{-	72	}$ & $	12.03	^{+	0.35	}_{-	0.51	}$ &	0.72	&$	6.23	^{+	0.22	}_{-	0.26	}$ & $	0.99	^{+	0.23	}_{-	0.22	}$ & $	11.192	^{+	0.046	}_{-	0.048	}$ & $	2.536	^{+	0.360	}_{-	0.338	}$\\
888.2533	&$	0.60	^{+	0.04	}_{-	0.04	}$ & $	586	^{+	37	}_{-	37	}$ & $	9.93	^{+	0.30	}_{-	0.38	}$ &	0.52	&$	6.30	^{+	0.19	}_{-	0.22	}$ & $	1.04	^{+	0.19	}_{-	0.17	}$ & $	8.085	^{+	0.037	}_{-	0.038	}$ & $	2.324	^{+	0.264	}_{-	0.248	}$\\
890.0729	&$	0.59	^{+	0.05	}_{-	0.04	}$ & $	605	^{+	36	}_{-	39	}$ & $	9.82	^{+	0.39	}_{-	0.36	}$ &	0.61	&$	6.43	^{+	0.19	}_{-	0.23	}$ & $	0.96	^{+	0.22	}_{-	0.20	}$ & $	7.560	^{+	0.035	}_{-	0.035	}$ & $	1.874	^{+	0.239	}_{-	0.222	}$\\
891.1884	&$	0.62	^{+	0.05	}_{-	0.04	}$ & $	596	^{+	37	}_{-	37	}$ & $	10.43	^{+	0.37	}_{-	0.35	}$ &	0.75	&$	6.16	^{+	0.20	}_{-	0.24	}$ & $	1.21	^{+	0.20	}_{-	0.18	}$ & $	7.825	^{+	0.035	}_{-	0.035	}$ & $	2.782	^{+	0.276	}_{-	0.262	}$\\
892.2633	&$	0.67	^{+	0.07	}_{-	0.05	}$ & $	566	^{+	48	}_{-	59	}$ & $	10.94	^{+	0.37	}_{-	0.39	}$ &	0.39	&$	5.80	^{+	0.24	}_{-	5.80	}$ & $	1.45	^{+	0.19	}_{-	0.18	}$ & $	7.619	^{+	0.035	}_{-	0.036	}$ & $	3.777	^{+	0.300	}_{-	0.290	}$\\
893.4686	&$	0.67	^{+	0.08	}_{-	0.06	}$ & $	570	^{+	51	}_{-	58	}$ & $	11.29	^{+	0.40	}_{-	0.43	}$ &	0.69	&$	6.06	^{+	0.27	}_{-	0.32	}$ & $	1.16	^{+	0.26	}_{-	0.23	}$ & $	7.265	^{+	0.034	}_{-	0.034	}$ & $	1.971	^{+	0.264	}_{-	0.249	}$\\
894.3870	&$	0.62	^{+	0.05	}_{-	0.05	}$ & $	542	^{+	42	}_{-	48	}$ & $	9.72	^{+	0.34	}_{-	0.35	}$ &	0.59	&$	6.34	^{+	0.23	}_{-	0.30	}$ & $	0.99	^{+	0.26	}_{-	0.23	}$ & $	5.417	^{+	0.026	}_{-	0.026	}$ & $	1.141	^{+	0.175	}_{-	0.161	}$\\
895.2340	&$	0.58	^{+	0.04	}_{-	0.04	}$ & $	590	^{+	37	}_{-	37	}$ & $	9.68	^{+	0.36	}_{-	0.36	}$ &	0.60	&$	6.38	^{+	0.25	}_{-	0.32	}$ & $	1.06	^{+	0.27	}_{-	0.24	}$ & $	5.619	^{+	0.026	}_{-	0.026	}$ & $	1.233	^{+	0.184	}_{-	0.170	}$\\
896.1426	&$	0.82	^{+	0.23	}_{-	0.10	}$ & $	487	^{+	70	}_{-	127	}$ & $	12.02	^{+	0.37	}_{-	0.21	}$ &	0.69	&$	6.43	^{+	0.25	}_{-	0.34	}$ & $	0.73	^{+	0.36	}_{-	0.30	}$ & $	5.939	^{+	0.027	}_{-	0.027	}$ & $	0.699	^{+	0.174	}_{-	0.153	}$\\
898.1645	&$	2.30	^{+	-0.28	}_{-	1.57	}$ & $	1	^{+	394	}_{-	1	}$ & $	9.50	^{+	0.15	}_{-	0.14	}$ &	0.92	&$	6.73	^{+	0.21	}_{-	0.25	}$ & $	0.70	^{+	0.30	}_{-	0.26	}$ & $	3.318	^{+	0.010	}_{-	0.010	}$ & $	0.389	^{+	0.050	}_{-	0.046	}$\\
899.0077	&$	0.74	^{+	0.19	}_{-	0.13	}$ & $	444	^{+	100	}_{-	137	}$ & $	9.92	^{+	0.38	}_{-	0.78	}$ &	1.09	&$	6.51	^{+	0.23	}_{-	0.31	}$ & $	0.84	^{+	0.34	}_{-	0.26	}$ & $	3.561	^{+	0.019	}_{-	0.019	}$ & $	0.580	^{+	0.111	}_{-	0.099	}$\\
901.2278	&$	2.90	^{+	-0.17	}_{-	2.15	}$ & $	1	^{+	416	}_{-	1	}$ & $	10.14	^{+	0.17	}_{-	0.11	}$ &	0.66	&$	6.10	^{+	0.30	}_{-	0.41	}$ & $	1.34	^{+	0.32	}_{-	0.26	}$ & $	2.852	^{+	0.007	}_{-	0.006	}$ & $	0.953	^{+	0.066	}_{-	0.064	}$\\
902.0756	&$	0.79	^{+	1.67	}_{-	0.12	}$ & $	407	^{+	90	}_{-	407	}$ & $	10.57	^{+	0.42	}_{-	0.52	}$ &	0.65	&$	6.54	^{+	0.23	}_{-	0.31	}$ & $	0.89	^{+	0.32	}_{-	0.26	}$ & $	2.408	^{+	0.013	}_{-	0.013	}$ & $	0.453	^{+	0.080	}_{-	0.073	}$\\
903.1209	&$	2.24	^{+	0.77	}_{-	1.35	}$ & $	23	^{+	320	}_{-	23	}$ & $	10.88	^{+	0.24	}_{-	0.22	}$ &	0.76	&$	6.77	^{+	0.20	}_{-	0.23	}$ & $	0.77	^{+	0.31	}_{-	0.26	}$ & $	1.550	^{+	0.005	}_{-	0.005	}$ & $	0.254	^{+	0.028	}_{-	0.027	}$\\
904.0978	&$	0.99	^{+	0.25	}_{-	0.23	}$ & $	199	^{+	194	}_{-	199	}$ & $	7.77	^{+	0.11	}_{-	0.10	}$ &	1.08	&$	6.03	^{+	0.24	}_{-	0.32	}$ & $	1.42	^{+	0.26	}_{-	0.21	}$ & $	1.037	^{+	0.004	}_{-	0.004	}$ & $	0.644	^{+	0.029	}_{-	0.029	}$\\
889.1765	&$	0.98	^{+	5.33	}_{-	0.14	}$ & $	443	^{+	85	}_{-	443	}$ & $	12.50	^{+	0.35	}_{-	0.36	}$ &	0.78	&$	5.83	^{+	0.26	}_{-	5.83	}$ & $	1.12	^{+	0.23	}_{-	0.20	}$ & $	9.245	^{+	0.042	}_{-	0.040	}$ & $	2.139	^{+	0.329	}_{-	0.312	}$\\
900.3845	&$	1.26	^{+	1.50	}_{-	0.54	}$ & $	142	^{+	302	}_{-	142	}$ & $	10.08	^{+	0.29	}_{-	0.29	}$ &	1.00	&$	6.15	^{+	0.33	}_{-	0.44	}$ & $	1.47	^{+	0.00	}_{-	0.27	}$ & $	2.751	^{+	0.015	}_{-	0.015	}$ & $	0.722	^{+	0.108	}_{-	0.100	}$\\
905.3416	&$	0.57	^{+	0.05	}_{-	0.05	}$ & $	291	^{+	30	}_{-	37	}$ & $	0.40	^{+	0.07	}_{-	0.06	}$ &	0.62	&$	5.60	^{+	0.18	}_{-	5.60	}$ & $	1.61	^{+	0.09	}_{-	0.14	}$ & $	0.778	^{+	0.006	}_{-	0.005	}$ & $	0.523	^{+	0.027	}_{-	0.030	}$\\
906.0508	&$	1.57	^{+	0.07	}_{-	0.07	}$ & $	672	^{+	70	}_{-	162	}$ & $	1.44	^{+	0.58	}_{-	0.18	}$ &	1.00	&$	5.60	^{+	0.15	}_{-	5.60	}$ & $	1.22	^{+	0.19	}_{-	0.19	}$ & $	0.686	^{+	0.002	}_{-	0.012	}$ & $	0.182	^{+	0.024	}_{-	0.024	}$\\
907.3148	&$	2.67	^{+	0.14	}_{-	0.14	}$ & $	684	^{+	28	}_{-	28	}$ & $	0.71	^{+	0.04	}_{-	0.04	}$ &	0.80	&$	5.65	^{+	0.15	}_{-	5.65	}$ & $	1.39	^{+	0.14	}_{-	0.14	}$ & $	0.476	^{+	0.004	}_{-	0.004	}$ & $	0.228	^{+	0.015	}_{-	0.015	}$\\
908.2906	&$	2.72	^{+	0.09	}_{-	0.26	}$ & $	823	^{+	29	}_{-	59	}$ & $	0.91	^{+	0.09	}_{-	0.04	}$ &	3.48	&$	5.60	^{	c	}_{	 	}$ & $	0.51	^{	c	}_{	 	}$ & $	0.358	^{+	0.002	}_{-	0.006	}$ & $	0.000	^{+	0.000	}_{-	0.000	}$\\
909.0731	&$	2.67	^{+	0.19	}_{-	0.31	}$ & $	697	^{+	56	}_{-	68	}$ & $	0.97	^{+	0.12	}_{-	0.04	}$ &	1.66	&$	6.79	^{	c	}_{	 	}$ & $	0.25	^{+	1.26	}_{-	0.25	}$ & $	0.253	^{+	0.003	}_{-	0.005	}$ & $	0.006	^{+	0.005	}_{-	0.005	}$\\
910.2218	&$	3.58	^{+	1.18	}_{-	0.87	}$ & $	668	^{+	111	}_{-	115	}$ & $	0.84	^{+	0.24	}_{-	0.22	}$ &	0.69	&$	6.41	^{	c	}_{-	0.38	}$ & $	0.25	^{+	1.20	}_{-	0.25	}$ & $	0.115	^{+	0.006	}_{-	0.005	}$ & $	0.011	^{+	0.006	}_{-	0.006	}$\\
904.1780	&$	1.66	^{	c	}_{-	0.94	}$ & $	1	^{+	387	}_{-	1	}$ & $	7.52	^{+	0.11	}_{-	0.11	}$ &	0.83	&$	5.73	^{+	0.17	}_{-	0.11	}$ & $	1.74	^{	c	}_{-	0.11	}$ & $	1.011	^{+	0.005	}_{-	0.006	}$ & $	0.911	^{+	0.039	}_{-	0.045	}$\\
904.3096	&$	0.80	^{+	0.79	}_{-	0.10	}$ & $	334	^{+	98	}_{-	327	}$ & $	7.76	^{+	0.14	}_{-	0.15	}$ &	0.75	&$	5.74	^{+	0.17	}_{-	0.10	}$ & $	1.75	^{	c	}_{-	0.11	}$ & $	0.988	^{+	0.005	}_{-	0.006	}$ & $	0.863	^{+	0.036	}_{-	0.042	}$\\
906.3936	&$	1.92	^{+	0.07	}_{-	0.08	}$ & $	736	^{+	91	}_{-	231	}$ & $	1.30	^{+	0.36	}_{-	0.22	}$ &	1.26	&$	5.60	^{+	0.21	}_{-	5.60	}$ & $	1.15	^{+	0.22	}_{-	0.21	}$ & $	0.640	^{+	0.002	}_{-	0.013	}$ & $	0.154	^{+	0.026	}_{-	0.026	}$\\
908.0061	&$	2.27	^{+	0.23	}_{-	0.18	}$ & $	778	^{+	68	}_{-	121	}$ & $	1.15	^{+	0.26	}_{-	0.13	}$ &	1.60	&$	5.60	^{	c	}_{	 	}$ & $	1.21	^{	c	}_{	 	}$ & $	0.399	^{+	0.007	}_{-	0.007	}$ & $	0.000	^{+	0.000	}_{-	0.000	}$\\
909.2486	&$	2.63	^{+	0.31	}_{-	0.23	}$ & $	683	^{+	82	}_{-	99	}$ & $	1.06	^{+	0.14	}_{-	0.14	}$ &	1.56	&$	6.02	^{	c	}_{	 	}$ & $	0.25	^{	c	}_{	 	}$ & $	0.238	^{+	0.005	}_{-	0.004	}$ & $	0.000	^{+	0.000	}_{-	0.000	}$\\
910.9905	&$	2.65	^{	c	}_{		}$ & $	529	^{+	770	}_{-	529	}$ & $	0.83	^{	c	}_{	 	}$ &	0.66	&$	7.00	^{	c	}_{	 	}$ & $	0.25	^{	c	}_{	 	}$ & $	0.069	^{+	0.016	}_{-	0.011	}$ & $	0.316	^{+	0.000	}_{-	0.000	}$\\
912.1792	&$	8.65	^{+	10.74	}_{-	5.72	}$ & $	1009	^{+	224	}_{-	538	}$ & $	1.17	^{+	0.30	}_{-	0.34	}$ &	0.86	&$	6.79	^{	c	}_{	 	}$ & $	1.70	^{	c	}_{	 	}$ & $	0.046	^{+	0.004	}_{-	0.005	}$ & $	0.000	^{+	0.000	}_{-	0.000	}$\\
911.0143	&$	4.50	^{+	5.39	}_{-	1.00	}$ & $	1011	^{+	288	}_{-	959	}$ & $	1.11	^{+	2.48	}_{-	0.14	}$ &	0.82	&$	5.94	^{	c	}_{	 	}$ & $	0.25	^{	c	}_{	 	}$ & $	0.076	^{+	0.006	}_{-	0.003	}$ & $	0.002	^{+	0.007	}_{-	0.000	}$\\
913.2270	&$	21.81	^{+	63.58	}_{-	18.50	}$ & $	1210	^{+	326	}_{-	687	}$ & $	1.63	^{	c	}_{	 	}$ &	0.54	&$	5.60	^{	c	}_{	 	}$ & $	1.49	^{	c	}_{	 	}$ & $	0.031	^{+	0.006	}_{-	0.006	}$ & $	0.000	^{+	0.000	}_{-	0.000	}$\\
914.1906	&$	13.72	^{+	57.73	}_{-	13.19	}$ & $	1070	^{	c	}_{		}$ & $	0.01	^{	c	}_{	 	}$ &	0.82	&$	6.64	^{	c	}_{	 	}$ & $	0.80	^{	c	}_{	 	}$ & $	0.017	^{+	0.006	}_{-	0.007	}$ & $	0.013	^{+	0.014	}_{-	0.011	}$\\
915.0509	&$	13.95	^{+	33.88	}_{-	11.33	}$ & $	1154	^{+	261	}_{-	430	}$ & $	0.98	^{	c	}_{	 	}$ &	0.77	&$	7.00	^{	c	}_{	 	}$ & $	0.25	^{	c	}_{	 	}$ & $	0.016	^{+	0.004	}_{-	0.004	}$ & $	0.004	^{+	0.010	}_{-	0.000	}$\\
916.1545	&$	87.75	^{+	306.47	}_{-	69.32	}$ & $	1195	^{+	467	}_{-	210	}$ & $	10.07	^{	c	}_{	 	}$ &	0.74	&$	5.60	^{+	0.90	}_{-	5.60	}$ & $	1.17	^{	c	}_{	 	}$ & $	0.020	^{+	0.004	}_{-	0.005	}$ & $	0.009	^{+	0.008	}_{-	0.008	}$\\
\hline
\end{tabular}
\end{center}
\footnotesize{ 
\begin{flushleft} $^{a}$ Unabsorbed flux of the EQPAIR model in units of $10^{-9}~{\rm erg~s^{-1}~cm^{-2}}$. \\
$^b$ Unabsorbed flux of the Gaussian component in units of $10^{-12}~{\rm erg~s^{-1}~cm^{-2}}$.\\
$^c$ Error bars could not be constrained either due to low flux level or degeneracy between model parameters. 
\end{flushleft} }
         \label{tab:par.2011-12}
\end{table*}

\label{lastpage}

\end{document}